\documentclass[a4paper,11pt]{article}

\usepackage{jheppub}
\usepackage{amsmath,amssymb,amsthm}
\usepackage{graphicx}
\usepackage{booktabs}
\usepackage{orcidlink}

\usepackage{comment}

	\def\be{\begin{equation}}
		\def\ee{\end{equation}}
	\def\ba{\begin{eqnarray}}
		\def\ea{\end{eqnarray}}

\definecolor{mmablue}{rgb}{0.368417,0.506779,0.709798}
\definecolor{mmaorange}{rgb}{0.880722,0.611041,0.142051}

\newcommand{\rs}{r_{s}}
\newcommand{\lL}{\lambda}
\newcommand{\eps}{\varepsilon}
\newcommand{\clm}{c_{\ell m}}
\newcommand{\dd}{\mathrm{d}}

	\def \black hole { {\bar h} }

	\def\IR{\relax{\rm I\kern-.18em R}}
	\def\IL{\relax{\rm I\kern-.18em L}}
	
	\def\inv{^{\raise.15ex\hbox{${\scriptscriptstyle -}$}\kern-.05em 1}}

	\def\bea{\begin{eqnarray}}
		\def\eea{\end{eqnarray}}

	\definecolor{markcolor2}{rgb}{1,0,0}
	
	\definecolor{markcolor3}{rgb}{0,1,0}

    \def\dd{{\rm d}}

\usepackage{graphics,appendix,afterpage,makecell} 

\definecolor{oucrimsonred}{rgb}{0.6, 0.0, 0.0}
\definecolor{persianblue}{rgb}{0.11, 0.22, 0.73}
\definecolor{forestgreen}{rgb}{0.13,0.35,0.13}
\definecolor{lightgray}{rgb}{0.83, 0.83, 0.83}
\definecolor{cornellred}{rgb}{0.7, 0.11, 0.11}
\definecolor{navyblue}{rgb}{0.0, 0.0, 0.5}
\definecolor{amethyst}{rgb}{0.6, 0.4, 0.8}
\definecolor{yellow}{rgb}{1.0, 1.0, 0.0}
\definecolor{firebrick}{rgb}{0.7, 0.13, 0.13}
\definecolor{tangerineyellow}{rgb}{1.0, 0.8, 0.0}
\definecolor{deepfuchsia}{rgb}{0.76, 0.33, 0.76}
\definecolor{amber}{rgb}{1.0, 0.75, 0.0}
\definecolor{VioletRed4}{rgb}{0.55, 0.13, .32}
\definecolor{indiagreen}{rgb}{0.07, 0.53, 0.03}
\definecolor{VioletRed4}{rgb}{0.55, 0.13, .32}

\definecolor{oucrimsonred}{rgb}{0.6, 0.0, 0.0}
\newcommand\vertarrowbox[3][6ex]{%
  \begin{array}[t]{@{}c@{}} #2 \\
  \left\uparrow\vcenter{\hrule height #1}\right.\kern-\nulldelimiterspace\\
  \makebox[0pt]{\scriptsize#3}
  \end{array}%
}
\hypersetup{
     colorlinks   = true,
     citecolor    = violet,
     urlcolor     = violet,
     linkcolor    = violet}

\definecolor{verdechiaro}{rgb}{0.6,1,0.6}
\definecolor{giallochiaro}{rgb}{1,1,0.6}
\definecolor{bluscuro}{rgb}{0.15, 0.2, 0.9}
\definecolor{verdes}{rgb}{0.1, 0.5, 0.1}%
\definecolor{tangerineyellow}{rgb}{1.0, 0.8, 0.0}

\definecolor{americanrose}{rgb}{1.0, 0.01, 0.24}
\definecolor{cobalt}{rgb}{0.0, 0.28, 0.67}
\definecolor{brandeisblue}{rgb}{0.0, 0.44, 1.0}
\definecolor{mycolor}{rgb}{0.0, 0.0, 0.5}
\definecolor{oxfordblue}{rgb}{0.0, 0.13, 0.28}
\definecolor{azure}{rgb}{0.0, 0.5, 1.0}
\definecolor{turquoiseblue}{rgb}{0.0, 1.0, 0.94}

\definecolor{verdes}{rgb}{0.1, 0.5, 0.1}%
\definecolor{cornellred}{rgb}{0.7, 0.11, 0.11}

\definecolor{VioletRed4}{rgb}{0.55, 0.13, .32}

\definecolor{rossocorsa}{rgb}{0.83, 0.0, 0.0}

\title{The Schwarzschild Black Hole
in an External Gravitational Tidal Field:  the Quasinormal Spectrum}

\author[a,b]{D.~Giataganas\orcidlink{0000-0003-2003-3902},}
\author[c]{A.~Kehagias\orcidlink{0000-0001-6080-6215},}
\author[d]{A.~Riotto\orcidlink{0000-0001-6948-0856}}

\affiliation[a]{Department of Physics, National Sun Yat-Sen University, Kaohsiung 80424, Taiwan
}
\affiliation[b]{Physics Division, National Center for Theoretical Sciences, Taipei 10617, Taiwan}
\affiliation[c]{Physics Division, National Technical University of Athens, Athens 15780, Greece}
\affiliation[d]{Department of Theoretical Physics and Gravitational Wave Science Center,  \\
24 quai E. Ansermet, CH-1211 Geneva 4, Switzerland}

\abstract{
A black hole need not ring down in isolation, since a nearby companion can subject it to an external gravitational tide. We calculate how a quadrupole tide modifies the Schwarzschild black hole resonances. Linearising the exact tidally distorted Schwarzschild solution in the tidal amplitude $A$, we derive the diagonal odd- and even-parity perturbation equations on the deformed background and compute the formal first order displacements of analytically continued Schwarzschild resonances. Each diagonal first order shift factorises as $\delta\omega_{n\ell m}=A\,c_{\ell m}\kappa_{n\ell}$, giving one reduced complex coefficient for each $(n,\ell)$, expressed as a ratio of contour integrals, multiplied by a universal Zeeman-like pattern that splits the $2\ell+1$ azimuthal multiplet. For the $\ell=2$ and $\ell=3$ sectors analysed explicitly, the odd- and even-parity shifts coincide, so Schwarzschild isospectrality survives at first order in the tidal amplitude. Quadratic tidal forces, by contrast, can break this degeneracy. Moreover, for a physical companion, the astrophysically dominant $(n,\ell,m)=(0,2,2)$ mode oscillates faster and decays more slowly. Finally, we show that the eikonal shifts admit a geometric description in terms of the Penrose limit about the tidally deformed photon ring.
}

\emailAdd{dimitrios.giataganas@gmail.com}
\emailAdd{kehagias@central.ntua.gr}
\emailAdd{Antonio.Riotto@unige.ch}
\begin{document}
\maketitle
\flushbottom

\section{Introduction}
General relativity is tested today using the gravitational waves emitted during the merger of two compact objects~\cite{LIGOScientific:2021sio}. Following the merger, the remnant rings down through its quasinormal modes~\cite{Kokkotas:1999bd,Berti:2009kk}. Black hole ringdown provides a particularly valuable probe of strong and dynamical gravity. Accurate predictions for its quasinormal frequencies and excitation amplitudes, confronted with gravitational-wave observations of compact-binary mergers, can sharpen our understanding of the two-body problem, test general relativity in its nonlinear regime, and search for additional gravitational degrees of freedom~\cite{Berti:2025hly}. A black hole, however, need not ring down in complete isolation. A nearby companion can generate an external gravitational tidal field, modifying the geometry in which the black hole oscillates~\cite{Poisson:2014gka}. This motivates the central problem addressed here: determining how such a tidal environment reshapes the quasinormal spectrum.

Much of the literature on environmental modifications of black-hole spectra
has focused on matter environments, dirty black holes surrounded by shells, halos, or clouds~\cite{Leung:1997was,Barausse:2014tra,Spieksma:2024voy,Bhowmik:2026owi,Zhao:2026eti}. In such settings, the quasinormal spectrum is deformed and the celebrated degeneracy between the odd- and even-parity spectra of Schwarzschild~\cite{Chandrasekhar:1985kt} can be broken~\cite{Li:2023ulk}. A binary companion, by contrast, places the black hole in a locally vacuum tidal environment: the companion produces external curvature but no local stress-energy in the neighbourhood of the black hole. The exact spacetime of a Schwarzschild black hole in a static quadrupolar tide is known~\cite{Kehagias:2024rtz} (for  Kerr, see Ref. \cite{Gounis:2024hcm}), and is used to compute its Love number, thus providing a natural laboratory in which the spectral deformation can be derived from first principles. 

In this paper, we carry out this program for gravitational spin-2 perturbations. We linearise the exact tidally distorted solution of
Ref.~\cite{Kehagias:2024rtz} in the tidal amplitude $A$, derive the diagonal $(\ell,m)$ perturbation equations in the odd-parity
Regge-Wheeler~\cite{Regge:1957td} and even-parity
Zerilli~\cite{Zerilli:1970se} sectors of the distorted background, and compute the first order shifts of the quasinormal frequencies using the bilinear form perturbation theory of Refs.~\cite{Lestingi:2026peq,Green:2022htq,Mark:2014aja}. Three structural results emerge. First, each diagonal first order shift factorises as
\begin{equation}
\delta\omega_{n\ell m}=A\,\clm\,\kappa_{n\ell}\,,\qquad \clm=\frac{\ell(\ell+1)-3m^{2}}{(2\ell-1)(2\ell+3)}\,,
\label{eq:master}
\end{equation}
where $A$ is the parameter controlling the tidal deformation. This   produces a Zeeman-like splitting of the $2\ell+1$ azimuthal multiplet. For an axisymmetric quadrupole, the relative splitting is fixed entirely by angular-momentum algebra, while its centroid remains unchanged:
$\sum_{m}\delta\omega_{n\ell m}=0$. Second, for the $\ell=2$ and $\ell=3$ multipoles considered explicitly, the odd- and even-parity spectra shift identically, contrary to the intuition developed from matter environments. For these multipoles, Chandrasekhar isospectrality therefore survives the tide at first order in the diagonal
sector. We trace this surviving degeneracy to an intertwining map that deforms smoothly with the tidal field.  Third, the reduced complex coefficient $\kappa_{n\ell}$ is expressed analytically as a ratio of contour integrals of closed-form densities constructed from the Regge-Wheeler mode function. For the astrophysically dominant  $(n,\ell,m)=(0,2,2)$ fundamental mode, common to both parity sectors, we find $\delta(M\omega)_{022}=-\big(1.43+0.75\,i\big)A$, where $M$ is the renormalised horizon mass scale. Thus, for a physical companion in our convention, $A<0$, this mode oscillates faster and decays more slowly. 

The  paper is organised as follows. Section~\ref{section:2} introduces the exact tidally distorted geometry, its linearisation in the tidal amplitude, and the coordinate and normalisation dictionary used in the wave calculation. Section~\ref{section:3} derives the diagonal odd- and even-parity perturbation equations and establishes their first-order intertwining relation. Section~\ref{section:4} formulates the contour-integral perturbation theory and evaluates the resulting quasinormal-frequency shifts. Section~\ref{section:5} develops the Penrose-limit description of the photon ring and derives the associated transverse oscillator spectrum. Section~\ref{section:6} obtains the same eikonal regime directly from the large-$\ell$ wave equations using the WKB approximation. Section~\ref{section:7} compares the geodesic and wave descriptions after accounting for their different normalisations. Section~\ref{section:8} summarises our conclusions and discusses possible extensions and phenomenological implications. The main text is supported by several Appendices. Appendix~\ref{App:A1} derives the linearised background, angular factor, and radial conventions; Appendix~\ref{App:A2} presents the odd-parity projection and reduction; Appendix~\ref{App:A3} gives the even-parity reduction, intertwiner, and Sturm--Liouville weight; and Appendix~\ref{App:A4} details the contour-integral evaluation and its numerical and analytical checks.

\section{The tidally distorted black hole}
\label{section:2}
Let us consider the exact non-perturbative solution of a Schwarzschild black hole immersed in an external
gravitational tidal field~\cite{Kehagias:2024rtz}. Restricting to a quadrupole tide, in the Schwarzschild-type coordinates $(t,r,\theta,\phi)$ in which   the horizon sits at
$r=\rs\equiv r_{s,\rm Weyl}=2M_{\rm Weyl}$, the metric reads
\begin{align}
\dd s^{2}=-\Big(1-\tfrac{\rs}{r}\Big)e^{\,2\mathcal U(r)P_{2}}\dd t^{2}
+e^{-2\mathcal U(r)P_{2}}
\Big[e^{2k_{0}(r,\theta)}\Big(\tfrac{\dd r^{2}}{1-\rs/r}+r^{2}\dd\theta^{2}\Big)
+r^{2}\sin^{2}\theta\,\dd\phi^{2}\Big],
\label{eq:metric}
\end{align}
with 
\begin{equation}
P_{2}=P_{2}(\cos\theta),\qquad
\mathcal U(r)\equiv A\bigg(\frac{4r^{2}}{\rs^{2}}-\frac{4r}{\rs}+\frac{2}{3}\bigg)
\end{equation}
and, by writing $y\equiv 2r/\rs-1$, we have 
\begin{equation}
k_{0}=-2A\,y\sin^{2}\theta+{\cal O}(A^{2}),\quad
\mathcal U(y)=A\left(y^2-\frac13\right)
\end{equation}
where the exact $k_{0}$ function is known but for our purposes only its linear part is needed.  This geometry belongs to the Weyl class of static axisymmetric vacuum solutions~\cite{papapetrou1953,Ernst:1967wx}. The branch present in our metric, growing as $y^{2}$, represents the applied quadrupolar field, whereas the decaying radial branch is singular at the horizon and is excluded by regularity. This absence of a regular decaying branch expresses the nonperturbative vanishing of the static tidal Love response~\cite{Kehagias:2024rtz,Combaluzier-Szteinsznaider:2024sgb}. Because the growing tidal field renders the exact geometry non-asymptotically flat, it is interpreted as the local near-zone background of the tidally perturbed black hole, to be matched to a global binary spacetime.

We can translate the abstract tidal parameter $A$ into physical binary parameters. For a companion of mass $M_c$ at separation $d$, choosing the polar axis to
point toward the companion, the large-separation limit gives, at leading
order,
\begin{equation}
A\simeq-\frac{M_cM_{\rm Weyl}^{2}}{d^{3}}=-\frac{M_c r_{s,\rm Weyl}^{2}}{4d^{3}},
\label{eq:matching}
\end{equation}
in the conventions used here. Thus
$|A|\sim{\cal E}M_{\rm Weyl}^{2}$, where
${\cal E}\sim M_c/d^{3}$ is the applied tidal-curvature scale.  As shown below, replacing $M_{\rm Weyl}$ by the renormalised horizon mass
scale changes this matching relation only at ${\cal O}(A^{2})$.
A gauge transformation constructed in Appendix~\ref{App:A1} brings the
quadrupolar sector of the ${\cal O}(A)$ metric to diagonal
Regge-Wheeler gauge.

The accompanying monopole sector is absorbed into the physical horizon
radius and the correspondingly normalised Killing time through
\begin{equation}
r_{s,\rm RW}=\left(1-\frac{2A}{3}\right)r_{s,\rm Weyl},
\qquad
t_{\rm RW}=\left(1+\frac{2A}{3}\right)t_{\rm Weyl}.
\label{eq:normalisation}
\end{equation}
Here $r_{s,\rm Weyl}$ is the horizon parameter and $t_{\rm Weyl}$ is the 
Killing time appearing in Eq.~\eqref{eq:metric}. Defining 
$M\equiv r_{s,\rm RW}/2$, the radius redefinition also implies
\begin{equation}
M=\left(1-\frac{2A}{3}\right)M_{\rm Weyl}.
\end{equation}
Consequently, replacing $M_{\rm Weyl}$ by $M$ in the matching relation \eqref{eq:matching} changes its right-hand side only at ${\cal O}(A^{2})$. 

Invariance of the mode
phase, $\omega_{\rm RW}t_{\rm RW}=\omega_{\rm Weyl}t_{\rm Weyl},$
then gives
\begin{equation}
\omega_{\rm RW}=\frac{\omega_{\rm Weyl}}{1+\frac{2A}{3}}=\left(1-\frac{2A}{3}\right)\omega_{\rm Weyl}+{\cal O}(A^2),
\end{equation}
and therefore
\begin{equation}
\frac{r_{s,\rm RW}~\omega_{\rm RW}}{r_{s,\rm Weyl}~\omega_{\rm Weyl}} 
=1-\frac{4A}{3}+{\cal O}(A^2).
\label{eq:frequencyconversion}
\end{equation}
Thus the horizon-radius redefinition contributes $-2A/3$, while the
inverse Killing-time normalisation contributes another $-2A/3$, producing
the total fractional conversion $-4A/3$ of the dimensionless frequency.

After these redefinitions, we drop the chart labels and write 
$r_s\equiv r_{s,\rm RW}=2M$, 
where $r_s$ is the physical areal horizon radius and $M$ is the corresponding
renormalised horizon mass scale. Likewise, $t\equiv t_{\rm RW}$ denotes the
renormalised Killing time. Setting from now on $r_s=1$, and hence $M=1/2$,
we define $f\equiv1-1/r$ and $P_2\equiv P_2(\cos\theta)$; 
the metric takes the form
\begin{align}
\dd s^{2}={}-f\big(1-\eps\,r^{2}fP_{2}\big)\dd t^{2}
+f^{-1}\big(1+\eps\,r^{2}fP_{2}\big)\dd r^{2}
+r^{2}\left[1+\eps\left(r^{2}-\tfrac12\right)P_{2}\right]
\dd\Omega^{2},
\label{eq:RWgauge}
\end{align}
with 
$\eps\equiv-8A$. 
Up to the overall tidal amplitude and trivial residual gauge
transformations, Eq.~\eqref{eq:RWgauge} is the unique growing static, even-parity
$\ell=2$ vacuum deformation regular at the horizon. In the Regge-Wheeler
decomposition used here, its radial amplitudes are
\begin{equation}
H_{0}=H_{2}=\eps r^{2}f,
\qquad
K=\eps\left(r^{2}-\tfrac12\right).
\end{equation}
Upon restoring $\rs=2M$, the angular amplitude has radial dependence
$K\propto r^{2}-2M^{2}$, corresponding to the horizon-regular tidal profile
of Binnington and Poisson~\cite{Binnington:2009bb}.

\section{Gravitational perturbations of the tidally distorted black hole}
\label{section:3}
 
\subsection{Odd parity: deformed Regge–Wheeler equation}
Let us consider a dynamical gravitational perturbation with time dependence
$e^{-i\omega t}$ on the background~\eqref{eq:RWgauge}. Angular momentum and
parity selection rules organise the computation. Because the tidal
background is an axisymmetric polar quadrupole with $(L,M)=(2,0)$,
angular-momentum addition allows a dynamical multipole $(\ell,m)$ to couple
at ${\cal O}(\eps)$ to
$\ell'=\ell,\ell\pm1,\ell\pm2$, while axisymmetry enforces $m'=m$.
Conservation of total parity further restricts couplings within a given polar
or axial sector to $\ell'=\ell,\ell\pm2$, whereas polar--axial couplings
occur for $\ell'=\ell\pm1$.

The perturbation is already diagonal in $m$, so the $(2\ell+1)$-fold
azimuthal degeneracy of the Schwarzschild spectrum requires no further
diagonalisation. In the absence of accidental degeneracies between modes
with different $(n,\ell)$, only the diagonal $\ell'=\ell$ coupling
contributes to the first order frequency shift. The $\ell'\neq\ell$
sidebands, including the opposite polar/axial components with
$\ell'=\ell\pm1$, enter the eigenfunction at ${\cal O}(\eps)$ but feed back
into the frequency only at ${\cal O}(\eps^{2})$.

Finally, although the unperturbed Schwarzschild problem is isospectral
between the axial and polar sectors at fixed $\ell$, these two modes have
opposite total parity and are not mixed by an even-parity quadrupolar
background. No additional degenerate diagonalisation is therefore required,
and the diagonal odd- and even-parity problems may be treated separately at
first order.

The tide transforms as the $T^{2}_{0}$ component of a spherical tensor.
The Wigner-Eckart theorem therefore fixes the $m$-dependence of every diagonal matrix element to be proportional to $\clm$ in
Eq.~\eqref{eq:master}; the dynamics determines only the reduced coefficient $\kappa_{n\ell}$. As shown below, this factorisation holds at the level of the radial operators themselves.

In the odd sector, the perturbation is described in Regge-Wheeler gauge by
two radial amplitudes $(h_{0},h_{1})$. Projecting the linearised vacuum
equations on the distorted background onto the axial harmonics, as detailed
in Appendix \ref{App:A2}, gives at ${\cal O}(\eps^{0})$ the classic Regge-Wheeler
system~\cite{Regge:1957td}. For example, the angular-tensor projection gives 
$i\omega h_{0}+f(fh_{1})'=0,$ while the ${\cal O}(\eps)$ terms deform both independent equations.
Eliminating $h_{0}$ perturbatively and introducing
$\psi=fh_{1}/r$, the diagonal $\ell=2$ equation reduces to
\begin{equation}
\big(f\psi'\big)'+\frac{\omega^{2}-V^{-}_2}{f}\,\psi
+\eps\,c_{2m}\Big[\frac{(r-1)(2r-3)}{r}\,\psi'
-\frac{2\omega^{2}r^{4}-15r^{2}+9r+4}{r^{2}}\,\psi\Big]=0\,,
\label{eq:oddmaster}
\end{equation}
where
\begin{equation}
V^{-}_{2}=f\left(\frac{6}{r^{2}}-\frac{3}{r^{3}}\right)
\end{equation}
is the $\ell=2$ Regge-Wheeler potential. Equation~\eqref{eq:oddmaster} is
the diagonal odd-parity master equation through ${\cal O}(\eps)$ relevant
to the first order spectral problem and, to our knowledge, has not appeared
previously.

The projection and reduction were performed independently for
$|m|=0,1,2$. The three deformation operators occur in the exact ratio 
$c_{20}:c_{21}:c_{22}=2:1:-2$, as can be seen from Eq. \eqref{eq:master}. Their proportionality to $c_{2m}$ shows that
the Wigner-Eckart factorisation is an operator identity and provides a
nontrivial check of the angular projections.

The part of the deformation proportional to $\omega^2$ fixes the leading radial characteristics and hence the local ingoing and outgoing boundary conditions.  As shown in Appendix \ref{App:A2}, writing a local solution as
$\psi\sim\exp[\int^{r}k(s)\dd s]$, the leading oscillatory part of the two large-$r$ characteristic roots can be written as
\begin{equation}
k_{\pm}(r)=\pm i\omega\,\frac{\dd x}{\dd r},
\qquad
\frac{\dd x}{\dd r}=\frac{1}{f}-\eps c_{2m}r^{2}+{\cal O}(\eps^{2}),
\end{equation}
where the mode-dependent radial phase function $x(r)$ is defined by the second equation and hence,
\begin{equation}
x=r_{*}-\eps c_{2m}\frac{r^{3}}{3}
+{\cal O}(\eps^{2}).
\label{eq:tortoise}
\end{equation}
Thus the local outgoing solution behaves as $e^{+i\omega x}$.
Algebraically, the diagonal equation requires
$|\eps c_{2m}|\,|r|^2\ll1$. The underlying linearised tidal geometry,
however, requires the stronger mode-independent condition
$|\eps|\,|r|^2\ll1$. We therefore restrict throughout to the overlap region
\begin{equation}
1\ll |r|\ll|\eps|^{-1/2}.
\label{eq:overlap}
\end{equation}
Below we show that the same characteristic coordinate is obtained in the even channel as described in the next section, so the diagonal odd- and even-parity problems possess the same local ingoing and outgoing boundary classes through ${\cal O}(\eps)$.

\subsection{Even parity: reduction and intertwining}

Having reduced the diagonal odd-parity problem to the deformed Regge-Wheeler equation~\eqref{eq:oddmaster}, we now carry out the corresponding reduction in the even sector. The central question is whether the resulting even-parity operator remains intertwined with the odd-parity operator at first order in the tidal deformation. In Regge-Wheeler gauge, the even-parity perturbation is described by four radial amplitudes $(H_{0},H_{1},H_{2},K)$ and yields seven projected Einstein equations, related by the linearised Bianchi identities. Their reduction is more involved than in
the odd sector but remains systematic. As detailed in Appendix \ref{App:A3}, the trace-free
angular equation determines $H_{0}$ algebraically in terms of the remaining
amplitudes, with $H_{0}=H_{2}+{\cal O}(\eps).$ Three further equations can then be solved for $(K',H_{1}',H_{2}')$. Substituting these expressions, together with the derivative of the trace-free relation, into the angular-trace equation produces an algebraic constraint that eliminates $H_{2}$. The remaining equations form the closed
first order system $u'=\left[\mathbb M_{0}(r,\omega)+\eps\,\clm\,\mathbb M_{1}(r,\omega)
\right]u,$ where $u=(K, H_{1})$ with the radial $2\times2$ coefficient matrices: the Schwarzschild coefficient matrix $\mathbb M_0$ and the diagonal first order tidal correction $\mathbb M_1$. The matrix entries are rational functions containing the usual Zerilli combination $\lL r+3M$, where $\lL=(\ell-1)(\ell+2)/2$. Eliminating $H_{1}$ gives
\begin{equation}
K''+p_{1}K'+p_{0}K=0,
\qquad
p_i=p_i^{(0)}+\eps\,\clm\,p_i^{(1)},
\end{equation}
with all coefficients given in closed form in Appendix \ref{App:A3}.

The comparison with the odd problem is most transparent if we work directly with $K$, rather than introducing the Zerilli master variable, for two reasons that also have a technical origin. First, for $\ell=2$, the undeformed $K$-equation is related to the
Regge-Wheeler equation by the rational first order differential map
\begin{equation}
K=-\frac{\omega^{2}r^{3}-12r^{2}-3r+3}{16\,r^{3}}\,\psi+\frac{(r-1)(4r+3)}{16\,r^{2}}\,\psi'.
\label{eq:intertwiner}
\end{equation}
For generic quasinormal frequencies, this map sends Regge-Wheeler solutions
to solutions of the $K$-equation while preserving their ingoing and outgoing
boundary classes. At $\eps=0$, it is the composition of the
Chandrasekhar--Detweiler transformation, which underlies Schwarzschild
isospectrality~\cite{Chandrasekhar:1985kt}, with the reconstruction of the
metric amplitude $K$ from the Zerilli master function. As shown in Appendix \ref{App:A3}, the
map admits a rational ${\cal O}(\eps)$ deformation that intertwines the
diagonal odd- and even-parity equations through first order in the tidal
field. Moreover, the principal part of the even equation gives the same mode-dependent characteristic coordinate as the odd equation \eqref{eq:tortoise},
\begin{equation}
x_{2m}^{\rm even}=x_{2m}^{\rm odd}=r_{*}-\eps c_{2m}\frac{r^{3}}{3}+{\cal O}(\eps^{2}).
\label{eq:even}
\end{equation}
Because the coefficients of the deformed intertwiner are rational and grow at most polynomially, they do not interchange the two exponential branches $e^{\pm i\omega x_{2m}}$. The map therefore preserves the local ingoing and outgoing boundary classes through ${\cal O}(\eps)$. This establishes the operator-level origin of the first order odd--even isospectrality found below.

Second, eliminating $H_{1}$ introduces apparent singularities at the three roots of
$q(r)\equiv \omega^{2}r^{3}+6r+3,$ although the original first order system and its solutions remain regular there. The coefficient of $K'$ in the undeformed equation can be written as
\begin{equation}
p_{1}^{(0)}=\frac{2}{r}+\frac{1}{r-1}-\frac{q'(r)}{q(r)}.
\end{equation}
Consequently, the integrating factor that puts the equation in
Sturm--Liouville form is the rational function
\begin{equation}
W(r)=\exp\left(\int p_{1}^{(0)}\,\dd r\right)=\frac{r^{2}(r-1)}{\omega^{2}r^{3}+6r+3}.
\label{eq:weight}
\end{equation}
Equivalently, the residues of $p_{1}^{(0)}$ are
$(2,1,-1,-1,-1)$ at $r=0$, $r=1$, and the three roots of $q(r)$.
Their sum vanishes, consistently with
$p_{1}^{(0)}={\cal O}(r^{-2})$ at infinity. Since $W$ is rational, these
isolated poles introduce no additional branch structure. The contour used
below is deformed, when necessary, in order to avoid them and the perturbation theory applies without further modification.

\section{Quasinormal-mode spectral shifts}
\label{section:4}
 
Having established that the diagonal odd- and even-parity problems are intertwined and possess the same boundary classes through ${\cal O}(\eps)$, we now compute their common first order spectral shift.
Quasinormal modes are resonances of an open system. Their radial mode
functions are not square integrable and, for the quasinormal boundary
conditions, grow exponentially toward both ends of the real radial axis.
Consequently, the usual Hermitian inner product is not available. The
perturbation theory appropriate to this problem~\cite{Lestingi:2026peq,Green:2022htq,Mark:2014aja} instead employs a bilinear form, without complex conjugation, on a complex radial Hankel-type contour ${\cal C}$ along which the analytically continued modes decay.

For an eigenvalue problem
$\hat L(\omega)\Psi+\eta\,\hat{\delta L}(\omega)\Psi=0,$
where $\eta$ denotes the relevant perturbative parameter, write
$\omega=\omega^{(0)}+\eta\omega^{(1)}+{\cal O}(\eta^2)$. The first order
solvability condition gives the analogue of the Rayleigh-Schrödinger
formula~\cite{Lestingi:2026peq},
\begin{equation}
\omega^{(1)}=-\frac{\big\langle\Psi,\delta\hat{ L}\Psi\big\rangle_{\cal C}}
{\big\langle\Psi,\partial_\omega\hat L\,\Psi\big\rangle_{\cal C}},
\qquad
\langle u,v\rangle_{\cal C}=\int_{\cal C}\dd r\,\mu(r,\omega)\,u\,v,
\label{eq:LSG}
\end{equation}
where $\mu$ is the integrating factor that makes the unperturbed radial
operator formally symmetric. Boundary terms vanish because the analytically
continued mode functions decay at the endpoints of ${\cal C}$.

In the odd sector, we use Eq.~\eqref{eq:oddmaster} directly in divergence
form, for which $\mu_{\rm odd}=1$. In the even sector, we use the monic
$K$-equation, for which $\mu_{\rm even}=W(r)$ given by Eq.~\eqref{eq:weight}.  The corresponding unperturbed even-parity eigenfunction is obtained from
Eq.~\eqref{eq:intertwiner}.

In units $\rs=1$, the unperturbed Schwarzschild mode is represented by the
Leaver series 
$\psi=e^{i\omega r}(r-1)^{-i\omega}r^{2i\omega}
\sum_{k=0}^{\infty}a_k
\left((r-1)/r\right)^k.$   Choosing the branch cut along the positive imaginary
direction makes its analytic continuation decay along the contour ${\cal C}$
~\cite{Leaver:1985ax,Lestingi:2026peq}.  
Specialising Eq.~\eqref{eq:master} to $\ell=2$, we define the reduced spectral-shift coefficient $\kappa_{n2}$   as $\delta\omega_{n2m}=A\,c_{2m}\kappa_{n2}.$
Applying Eq.~\eqref{eq:LSG} with $\eta=\eps c_{2m}$, and using $\eps=-8A$, gives
\begin{equation}
\kappa_{n2}=\frac{8\displaystyle\int_{\cal C}\dd r\,\psi
\Big[(r-1)(2r-3)\frac{\psi'}{r}
-\left(2\omega^{2}r^{4}-15r^{2}+9r+4\right)\frac{\psi}{r^2}\Big]}
{\displaystyle 2\omega\int_{\cal C}\dd r\,\frac{\psi^{2}}{f}}\,.
\label{eq:kappaint}
\end{equation}
Here $\psi$ and $\omega$ are the unperturbed Schwarzschild quantities. The
factor $1/f$ in the denominator comes from
$\partial_\omega\hat L_0=2\omega/f$, while the overall factor $8$ follows
from $\eps=-8A$. The analogous $\ell=3$ expression is obtained using the
operator given in Appendix \ref{App:A2}.

Because the applied quadrupolar field grows as $r^2$, the exact tidal
geometry is not asymptotically flat and does not supply a standard outgoing
boundary condition at infinity.   Equation~\eqref{eq:kappaint} is therefore
interpreted as the first-order displacement of the analytically continued
Schwarzschild resonance. A particular global binary completion can in principle modify the pole spectrum~\cite{Jaramillo:2020tuu,Cheung:2021bol}; the perturbative displacement itself, however, is well defined by analytic continuation. This construction is analogous in spirit to the
Stark problem, where an external field destroys ordinary bound states but
their continuations as resonance poles possess well-defined asymptotic
perturbative expansions~\cite{Graffi:1978gp,Herbst:1979dil}.

The contour ${\cal C}$ implements the corresponding continuation here.
Writing $\omega=\omega_R-i\Gamma$, with $\Gamma>0$, the outgoing mode behaves
along each vertical leg as $|\psi|\sim\exp\!\left[\Gamma\,\operatorname{Re}(r)- \omega_R\,\operatorname{Im}(r)\right],$
up to powers of $r$. Along these legs, $\operatorname{Re}(r)$ remains bounded
while $\operatorname{Im}(r)\to+\infty$, so the bilinear integrands decay as
$e^{-2\omega_R\operatorname{Im}(r)}$ times polynomial factors. The contour
integrals therefore remain finite despite the radial growth of the tidal
operator. The detailed contour construction and its convergence are
described in Appendix~\ref{App:A4}.

Evaluating Eq.~\eqref{eq:kappaint} and its even-parity counterpart gives, for
the $\ell=2$ fundamental mode and first overtone,
\begin{equation}
\kappa_{02}=9.98+5.28\,i,
\qquad
\kappa_{12}=1.85+10.56\,i.
\label{eq:kappavals}
\end{equation}
Applying the corresponding procedure at $\ell=3$ gives
\begin{equation}
\kappa_{03}=13.64+4.45\,i.
\label{eq:kappal3}
\end{equation}
The $\ell=3$ odd-parity deformation operator contains no term proportional
to $\omega^{2}$ and therefore produces no ${\cal O}(\eps r^{3})$
correction to its local characteristic phase, as shown in Appendix \ref{App:A2}.

For every multiplet computed here, the odd- and even-parity coefficients coincide. Apart from their common unperturbed Schwarzschild mode, the two calculations use different projected systems, radial variables, perturbing operators, and bilinear measures. Their numerical agreement is also checked by direct integration of the equations truncated at ${\cal O}(\eps)$ for several small finite values of $\eps$ in Appendix \ref{App:A4}. The exact equality through first order follows from the rational ${\cal O}(\eps)$ deformation of the intertwining map constructed in Appendix \ref{App:A3}.

\begin{figure}[t]
    \centering
\includegraphics[width=0.99\linewidth]{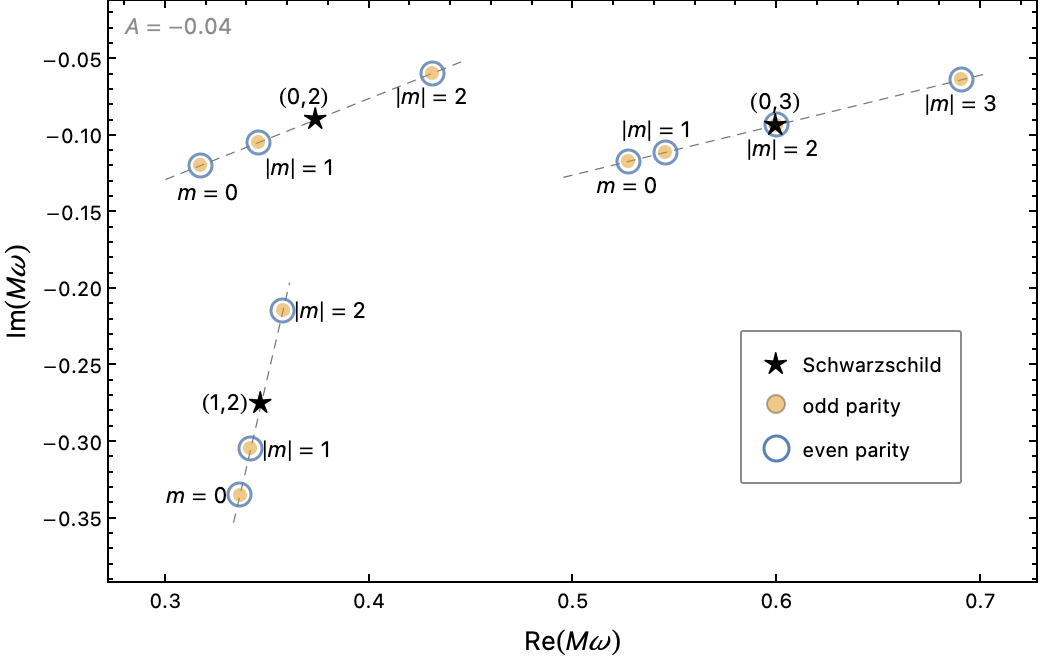}
    \caption{Tidal Zeeman splitting of the $(n,\ell)=(0,2)$, $(1,2)$, and $(0,3)$ quasinormal multiplets in the complex-frequency plane. Black stars mark the unperturbed Schwarzschild frequencies, while filled gold and open blue circles denote the odd- and even-parity modes, respectively. Their overlap illustrates the first order parity isospectrality. For each fixed $(n,\ell)$, the different $|m|$ components lie on a straight dashed line because $\delta\omega_{n\ell m}=A c_{\ell m}\kappa_{n\ell}$ with real $c_{\ell m}$; the $m$ and $-m$ modes remain degenerate. For $\ell=3$, the $|m|=2$ modes remain at the Schwarzschild value because $c_{32}=0$. For visibility, the linearised shifts are shown at the amplitude $A=-0.04$.
}
    \label{plot}
\end{figure}

The static quadrupolar vacuum tide therefore splits each azimuthal multiplet while preserving the odd-even parity doublet through ${\cal O}(A)$, in contrast with matter environments, where Schwarzschild isospectrality can be
broken~\cite{Li:2023ulk}. As shown in Fig.~\ref{plot}, every multiplet fans out along a straight line in the complex-frequency plane, whose direction is
set by $\kappa_{n\ell}$ and rotates between the fundamental mode and the
first overtone. At $\ell=3$, the quadrupolar selection rule gives
$c_{32}=0$, so the $|m|=2$ modes are unchanged at first order. The opposite axial/polar $\ell\pm1$ sidebands enter the eigenfunctions at ${\cal O}(A)$ but feed back into the diagonal frequency only at ${\cal O}(A^{2})$. We remark therefore that the parity degeneracy is  not guaranteed to
survive at second order.

For the $m=2$ member of the dominant $(n,\ell)=(0,2)$ fundamental
multiplet, with $m$ defined relative to the tidal symmetry axis, one has
$c_{22}=-2/7$. Restoring dimensions with $M=\rs/2$, Eq.~\eqref{eq:master} gives
\begin{equation}
\delta(M\omega)_{022}=\frac{A}{2}\,c_{22}\,\kappa_{02}
=-\frac{A}{7}\,\kappa_{02}=-\bigl(1.43+0.75\,i\bigr)A,
\label{eq:result22}
\end{equation}
for both parity sectors. Here $\omega$ is measured with respect to the RW-normalised Killing time
$t_{\rm RW}$, while $M=r_{s,\rm RW}/2$ is the renormalised horizon mass
scale of the
normalised Regge-Wheeler metric~\eqref{eq:RWgauge}. Using the Schwarzschild
value $M\omega_{022}^{(0)}=0.374-0.089\,i,$ the fractional changes in the oscillation frequency and damping rate are
\begin{equation}
\frac{\delta\omega_R}{\omega_R}\simeq-3.82\,A,
\qquad
\frac{\delta\gamma}{\gamma}\simeq+8.49\,A,
\qquad
\gamma\equiv-{\rm Im}\,\omega.
\label{eq:fractional}
\end{equation}
Equivalently, for the damping time $\tau=1/\gamma$,
\begin{equation}
\frac{\delta\tau}{\tau}\simeq-8.49\,A.
\end{equation}
Thus the fractional response of the damping time is more than twice that of
the oscillation frequency.

For a physical companion, $A<0$. The $m=\pm2$ doublet therefore shifts to a
higher oscillation frequency and a smaller damping rate, and hence rings
higher and longer. The $m=0$ singlet and the $m=\pm1$ doublet shift in the
opposite direction, with $c_{20}:c_{21}:c_{22}=2:1:-2.$
Including the $\pm m$ multiplicities, $c_{20}+2c_{21}+2c_{22}=0$. It follows that
$\sum_{m=-2}^{2}\delta\omega_{02m}=0$: the multiplet splits, but its
unweighted centroid in the complex-frequency plane remains fixed through
first order.

\section{The Penrose limit}
\label{section:5}
 
The eikonal regime of the spectrum computed above admits a purely geometric
description. Consider a null geodesic $\gamma$ with affine parameter $u$ and
tangent $k^\mu$. Its Penrose limit~\cite{Penrose1976,Blau:2003dz} is the plane
wave
\begin{equation}
\dd s_{\cal P}^{2}=2\,\dd u\,\dd V+{\cal A}_{ab}X^{a}X^{b}\dd u^{2}
+\dd X^{a}\dd X_{a}\,.
\label{eq:brinkmann}
\end{equation}
The transverse profile is determined by the curvature evaluated along
$\gamma$,
\begin{equation}
{\cal A}_{ab}=-R_{\mu\alpha\nu\beta}e^{\mu}_{a}k^{\alpha}
e^{\nu}_{b}k^{\beta}\big|_{\gamma},
\qquad
\ddot X^{a}={\cal A}^{a}{}_{b}X^{b},
\label{eq:deviation}
\end{equation}
where a dot denotes differentiation with respect to $u$. Completing $k^\mu$
to a null frame satisfying $k\cdot\ell=-1$ and
$e_a\cdot e_b=\delta_{ab}$, the inverse metric takes the form
$g^{\alpha\beta}=-k^{\alpha}\ell^{\beta}-\ell^{\alpha}k^{\beta}
+\delta^{ab}e^{\alpha}_{a}e^{\beta}_{b}.$ 
The Riemann symmetries then give
\begin{equation}
0=R_{\mu\nu}k^{\mu}k^{\nu}=\delta^{ab}R_{akbk}=-\,{\rm tr}\,{\cal A}
\label{eq:trace}
\end{equation}
on a Ricci-flat background. Since ${\cal A}_{ab}$ is a real symmetric
two-dimensional matrix, its instantaneous eigenvalues therefore have equal
magnitude and opposite sign,
\begin{equation}
{\rm Eig}({\cal A})=\big\{+\sigma^{2}(u),-\sigma^{2}(u)\big\}.
\label{eq:Adiag}
\end{equation}
For the equatorial circular photon orbit considered below, stationarity and
equatorial reflection symmetry make the radial and polar directions fixed
eigendirections and make $\sigma$ constant. The positive eigenvalue describes
the radial instability, while the negative one describes stable polar
oscillations. These transverse rates determine the imaginary and angular
parts of the eikonal spectrum~\cite{Cardoso:2008bp,Giataganas:2024hil}.

We now evaluate them directly in the coordinates of
Eq.~\eqref{eq:metric}, with $r_{s,\rm Weyl}=1$. On the equatorial plane,
$P_2(0)=-1/2$, and  
the squared impact parameter of a circular null orbit is therefore
\begin{equation}
b^{2}(r) =\frac{g_{\phi\phi}}{-g_{tt}}=\frac{r^{3}}{r-1}\,e^{2{\cal U}(r)}.
\end{equation}
Extremising it and selecting the root continuously connected to the Schwarzschild photon sphere gives 
\begin{equation}
r_{c}=\frac32(1-4A)+{\cal O}(A^{2}).
\end{equation}
This is a coordinate radius in the original Weyl-type chart and coincides with the corresponding  result of~\cite{Cardoso:2021qqu} after the coordinate and normalisation conversion discussed below in Eq. \eqref{eq:comparison}. 

To determine the radial instability, let
\begin{equation}
N\equiv(\ln b^{2})''=\frac{2r-1}{r^{2}(r-1)^{2}}-\frac{2}{r^{2}}+16A.
\end{equation}
Expanding the radial null-geodesic equation about $r=r_c$ gives
$\dd\delta r/\dd t=\lambda_L\delta r$ when the orbit is parametrised by the Killing
time. Equivalently,
\begin{equation}
\lambda_{L}^{2}(A)=\frac{N(r_c)}{2}\Big(1-\frac{1}{r_c}\Big)^{2}e^{-2{\cal U}(r_c)-2k_{0}(r_c,\frac{\pi}{2})}.
\label{eq:lamfull}
\end{equation}
Using the linearised expressions for ${\cal U}$ and $k_0$ yields
\begin{equation}
\lambda_{L}^{2} = \frac{4}{27}+\frac{272}{81}A,
\qquad
\lambda_{L} = \frac{2}{3\sqrt3} \left(1+\frac{34}{3}A\right).
\label{eq:lamclosed}
\end{equation}

The second transverse direction describes small departures from the
equatorial plane. Writing $\theta=\pi/2+\delta\theta$ and holding the
circular-orbit impact parameter $b_c=L/E$ fixed, the linearised polar
geodesic equation is
\begin{equation}
\ddot{\delta\theta}
+\Omega_{\rm aff}^{2}\delta\theta=0,
\qquad
\Omega_{\rm aff}^{2} =\frac{g^{\theta\theta}}{2}
\partial_{\theta}^{2} \left(g^{tt}+b_c^{2}g^{\phi\phi}\right)
E^{2}\bigg|_{\substack{r=r_c\\ \theta=\frac{\pi}{2}}}.
\label{eq:vert}
\end{equation}
The corresponding Killing-time frequency is
\begin{equation}
\Omega_{\theta} = \frac{\Omega_{\rm aff}}{|\dot t|}.
\end{equation}
Because $\dot t$ is constant along the circular orbit, the affine parameter
may be rescaled so that $\dot t=1$, and hence $u=t$ on the orbit. The radial
and polar directions are respectively the positive and negative
eigendirections of the traceless matrix ${\cal A}_{ab}$. Their rates must therefore have equal magnitude,
\begin{equation}
\Omega_{\theta}(A)=\lambda_L(A).
\label{eq:vertlyap}
\end{equation}
This equality is exact for the equatorial circular photon orbit of the Ricci-flat tidal geometry and is also verified directly from the exact metric.

The azimuthal orbital frequency is not an eigenvalue of the transverse
Penrose profile. It instead describes motion along the reference ray,
$\Omega_\phi=\dd\phi/\dd t$, and is given by
\begin{equation}
\Omega_{\phi}^{2} = \frac{4}{27}-\frac{88}{81}A.
\label{eq:Omphi}
\end{equation}
Thus $\Omega_\phi$ shifts oppositely to the two transverse rates. As a consistency check we note that in the Schwarzschild limit $(A=0)$, all three coincide; restoring the horizon radius,
\begin{equation}
\lambda_L=\Omega_\theta=\Omega_\phi=\frac{2}{3\sqrt3\,r_{s,\rm Weyl}}.
\label{eq:triple}
\end{equation}
The resulting triple degeneracy is what makes the eikonal spectrum of an isolated
black hole depend on a single number. 
The tide preserves the equality
$\lambda_L=\Omega_\theta$, which follows from the tracelessness of the
two-dimensional transverse curvature matrix, but breaks the equality with
$\Omega_\phi$, which has no corresponding Ricci-flatness protection.

\subsection{Oscillators and the eikonal spectrum}
 
Let us now promote the two transverse geodesic channels to waves. A minimally
coupled massless scalar field on the plane wave~\eqref{eq:brinkmann}
obeys~\cite{Fransen:2023eqj,Giataganas:2024hil}
\begin{equation}
\Box\Psi=2\,\partial_{u}\partial_{V}\Psi-{\cal A}_{ab}X^{a}X^{b}\partial_{V}^{2}\Psi
+\delta^{ab}\partial_{a}\partial_{b}\Psi=0.
\label{eq:box}
\end{equation}
Although we use a scalar field as a representative, the principal part of
the field equations is universal for massless fields in the eikonal limit:
their rapidly varying phase propagates along the same null rays.
Spin-dependent curvature couplings affect the amplitude transport and enter
the spectrum only at subleading order in the eikonal expansion. The vector $\partial_V$ is covariantly constant and Killing, so its
conjugate momentum $p$ is conserved. Taking $p>0$, we use the separation
ansatz
\begin{equation}
\Psi=e^{\,ipV-iEu}\,\psi_{r}(X_{r})\psi_{\perp}(X_{\perp}),
\qquad
E=E_{r}+E_{\perp}.
\label{eq:ansatz}
\end{equation}
In the eigenbasis
${\cal A}_{ab}={\rm diag}\big(\lambda_L^{2},-\Omega_\theta^{2}\big),$
the wave equation becomes
\begin{equation}
\left[2pE+\partial_{X_r}^{2}+\partial_{X_\perp}^{2}+p^{2}\lambda_L^{2}X_r^{2}-p^{2}\Omega_\theta^{2}X_\perp^{2}\right]\psi_r\psi_\perp=0.
\end{equation}
It therefore separates into a pair of Schr\"odinger problems with effective
mass $p$:
\begin{align}
\left[-\frac{1}{2p}\frac{\dd^{2}}{\dd X_{\perp}^{2}}
+\frac{p}{2}\Omega_{\theta}^{2}X_{\perp}^{2}\right]\psi_{\perp}
&=E_{\perp}\psi_{\perp},
\label{eq:normal}\\[2pt]
\left[-\frac{1}{2p}\frac{\dd^{2}}{\dd X_{r}^{2}}-\frac{p}{2}\lambda_{L}^{2}X_{r}^{2}\right]\psi_{r}
&=E_{r}\psi_{r}.
\label{eq:inverted}
\end{align}

Equation~\eqref{eq:normal} is the ordinary harmonic oscillator. Its
normalisable solutions are
$\psi_{\perp,j}
\propto
H_j\!\left(\sqrt{p\Omega_\theta}\,X_\perp\right)
e^{-\frac12p\Omega_\theta X_\perp^{2}},$
with the real ladder
\begin{equation}
E_{\perp}=\left(j+\frac12\right)\Omega_\theta,
\qquad
j=0,1,2,\ldots.
\label{eq:ladder}
\end{equation}
Equation~\eqref{eq:inverted} is the corresponding inverted oscillator. 
Outgoing conditions at $X_r\to\pm\infty$ select its resonant states. They are obtained by analytically continuing the oscillator frequency according to
$\Omega_\theta\to-i\lambda_L$: 
$\psi_{r,n}
\propto
H_n\!\left(e^{-i\pi/4}\sqrt{p\lambda_L}\,X_r\right)
e^{+\frac{i}{2}p\lambda_LX_r^{2}},$ 
with complex resonance energies
\begin{equation}
E_r =-i\left(n+\frac12\right)\lambda_L,
\qquad
n=0,1,2,\ldots.
\label{eq:resonances}
\end{equation}
The phase $p\lambda_LX_r^{2}/2$ has radial momentum
$p\lambda_LX_r$, and hence describes flux moving outward at both
$X_r\to+\infty$ and $X_r\to-\infty$. The half-integer offset in
Eq.~\eqref{eq:resonances} is the analytic continuation of the oscillator
zero-point term; the inverted oscillator itself has no normalisable ground
state.

It remains to relate the separation energy $E$ to the spacetime frequency.
With $\dd t/\dd u=1$, one has $u=t$ along the circular ray and
$\dd\phi/\dd u=\Omega_\phi^{\rm signed}$. The phase of a mode
$e^{-i\omega t+im\phi}$ restricted to the ray is therefore 
$-i\omega t+im\phi
= -i\left(\omega-m\Omega_\phi^{\rm signed}\right)u.$ 
Choosing the orientation of the ray associated with the sign of $m$ gives
$m\Omega_\phi^{\rm signed}=|m|\Omega_\phi$, and hence 
$E=\omega-|m|\Omega_\phi.$ 
For the near-equatorial eikonal family, with
$j=\ell-|m|={\cal O}(1)$ and fixed $n$ as $\ell\to\infty$, the three
contributions consequently assemble~\cite{Yang:2012he,Cardoso:2008bp} into
\begin{equation}
\omega \simeq |m|\Omega_\phi +\left(j+\frac12\right)\Omega_\theta
-i\left(n+\frac12\right)\lambda_L +{\cal O}(\ell^{-1}).
\label{eq:penrosespec}
\end{equation}
In the Weyl normalisation of Eq.~\eqref{eq:metric},
\begin{equation}
\Omega_\phi=\Omega_0\left(1-\frac{11}{3}A\right),
\qquad
\Omega_\theta=\lambda_L=\Omega_0\left(1+\frac{34}{3}A\right),
\end{equation}
with $\Omega_0=2/(3\sqrt3\,r_{s,\rm Weyl}).$
Thus the two transverse level spacings remain exactly equal, while their
common scale separates from the longitudinal orbital frequency according to
\begin{equation}
\frac{\Omega_\theta}{\Omega_\phi}=\frac{\lambda_L}{\Omega_\phi}
=1+15A+{\cal O}(A^{2}).
\label{eq:ratesplit}
\end{equation}
This ratio is unaffected by the common horizon-radius and Killing-time
normalisation discussed below.

\section{The eikonal limit of the wave computation}
\label{section:6}
 
The same regime follows directly from the large-$\ell$ limit of the projected
wave equations. After the Liouville transformation to normal form, the
leading eikonal equation is
\begin{equation}
\frac{\dd^{2}\psi}{\dd x^{2}}+\big(\omega^{2}-\widetilde V\big)\psi=0,
\end{equation}
with
\begin{align}
\widetilde V
&=\Lambda\big(g_0+q\,g_1\big)+{\cal O}(\Lambda^{0}),
\quad
q\equiv\eps\clm,
\quad
\Lambda=\ell(\ell+1),
\nonumber\\
g_0
&=\frac{f}{r^{2}},
\qquad
g_1=-\frac{f(4r^{2}-2r-1)}{2r^{2}}.
\label{eq:eikpot}
\end{align}
The ${\cal O}(\Lambda)$ potential is common to all massless fields.
The Liouville transformation also defines a mode-dependent normal form radial variable $x$ by requiring the principal wave operator to take the canonical form $\partial_x^2+\omega^2$. At leading eikonal order,
\begin{equation}
x=r_*+q\,\frac{r^3}{3}+{\cal O}(q^2,\Lambda^{-1}).
\label{eq:eikx}
\end{equation}
This leading-eikonal coordinate need not coincide with the finite-$\ell$ gravitational characteristic coordinate in Eqs.~\eqref{eq:tortoise} and~\eqref{eq:even} derived in Appendices \ref{App:A2} and \ref{App:A3}; in particular, their first order corrections have opposite signs at $\ell=2$.

For fixed overtone number as $\ell\to\infty$, the leading Schutz--Will
condition~\cite{Schutz:1985km} is
\begin{equation}
\omega^{2}\simeq\widetilde V(r_{\rm p})-i\left(n+\frac12\right)\sqrt{-2\frac{\dd^{2}\widetilde V}{\dd x^{2}}}\Bigg|_{r=r_{\rm p}}
,
\quad
\frac{\dd\widetilde V}{\dd r}\Bigg|_{r=r_{\rm p}}=0.
\label{eq:SW}
\end{equation}
We denote the maximum of the deformed eikonal potential by $r_p$. 
In the units $r_{s,\rm RW}=1$, the unperturbed peak lies at $r_0=3M=3/2$, where  $g_0(r_0)=4/27$ and  $g_1(r_0)=-10/27.$  Writing $r_{\rm p}=r_0+q\,r_1$, the peak condition gives
\begin{equation}
r_1=-\frac{g_1'(r_0)}{g_0''(r_0)}=-\frac{15}{8},
\qquad
r_{\rm p}=\frac32-\frac{15}{8}q+{\cal O}(q^2).
\end{equation}
Since $g_0'(r_0)=0$, the displacement of the peak does not contribute to its
height at first order, and one finds
\begin{equation}
\widetilde V(r_{\rm p})=\frac{4\Lambda}{27}\left(1-\frac52q\right)+{\cal O}(q^2,\Lambda^0).
\label{eq:Vpeak}
\end{equation}
The curvature of the potential does depend both on the displacement of the
peak and on the deformation of the characteristic coordinate
\eqref{eq:eikx}. Including both effects gives
\begin{equation}
\left.\frac{\dd^{2}\widetilde V}{\dd x^{2}}\right|_{r=r_{\rm p}}
=
-\frac{32\Lambda}{729}\left(1+\frac52q\right)+{\cal O}(q^2,\Lambda^0).
\label{eq:Vcurv}
\end{equation}

Writing $\omega=\omega_R+i\omega_I$, with $\omega_I<0$, substituting \eqref{eq:Vpeak} and \eqref{eq:Vcurv} in
Eq.~\eqref{eq:SW} and expanding gives
\begin{align}
\omega_R
&=\frac{2\sqrt{\Lambda}}{3\sqrt3}\left(1-\frac54q\right)+{\cal O}(\Lambda^{-1/2},q^2),
\nonumber\\
\omega_I&=-\left(n+\frac12\right)\frac{2}{3\sqrt3}\left(1+\frac52q\right)+{\cal O}(\Lambda^{-1},q^2).
\label{eq:eikfreq}
\end{align}
The first order shifts are therefore
\begin{align}
\delta\omega_R=-\eps\clm\frac{5\sqrt3}{18}\sqrt{\Lambda},\quad 
\delta\omega_I= -\eps\clm\frac{5\sqrt3}{18}(2n+1).
\label{eq:eikshifts}
\end{align}
Finally, using $\eps=-8A$ and
$\sqrt{\ell(\ell+1)}=\ell+\tfrac12+{\cal O}(\ell^{-1})$, these shifts take
the factorised form of Eq.~\eqref{eq:master}, with
\begin{equation}
\kappa^{\rm eik}_{n\ell}=\frac{20\sqrt3}{9}
\left[\left(\ell+\frac12\right)+i(2n+1)\right]+{\cal O}(\ell^{-1}).
\label{eq:kappaeik}
\end{equation}

The leading eikonal response is independent of the spin of the perturbing
field. At finite $\ell$, however, the scalar and gravitational shifts differ
substantially. For example, for the fundamental $\ell=2$ mode,
$\kappa^{s=0}_{02}=7.20+3.02\,i,$ $\kappa^{s=2}_{02}\equiv\kappa_{02}=9.98+5.28\,i.$ 
Their characteristic radial phases also receive corrections of opposite sign
at $\ell=2$, while the gravitational equation at $\ell=3$ has no correction at this order to its characteristic phase (Appendices~\ref{App:A2} and~\ref{App:A4}). These finite-$\ell$ differences are subleading in the eikonal expansion and are consistent with the common limit
~\eqref{eq:kappaeik}. In particular, ${\rm Re}\,\kappa/(\ell+\tfrac12)$ approaches
$20\sqrt3/9\simeq3.85$ from below in the scalar case and from above in the gravitational case, while
${\rm Im}\,\kappa/(2n+1)$ approaches the same limit.

The geometric origin of this convergence is transparent in the Penrose
description. At leading eikonal order, every massless field propagates along
the same null rays, and its transverse dynamics is controlled by the
curvature matrix ${\cal A}_{ab}$ of Eq.~\eqref{eq:brinkmann}. The leading
${\cal O}(\ell)$ real shift and ${\cal O}(1)$ damping shift are therefore
properties of the geometry alone. Dependence on the spin and detailed wave
potential enters only through subleading WKB orders.

\section{Geodesic interpretation of the wave spectrum}
\label{section:7}

The Penrose-limit and WKB calculations provide complementary descriptions of the same eikonal physics, with a direct mode-by-mode comparison in the near-equatorial sector. The former decomposes the spectrum into the azimuthal orbital frequency $\Omega_\phi$, the stable polar frequency $\Omega_\theta$, and the radial instability rate $\lambda_L$, whereas the latter determines their combined imprint on the quasinormal frequency through $\kappa^{\rm eik}_{n\ell}$. A direct comparison is slightly obscured because the two calculations use different horizon-radius and Killing-time normalisations. We first establish the corresponding dictionary and then show agreement both for the individual photon ring rates and for the full near-equatorial spectrum.

The wave calculation uses the physical horizon radius and Killing time defined in Eq.~\eqref{eq:normalisation} and derived in Appendix \ref{App:A1}, whereas the geodesic calculation uses the Weyl-chart quantities appearing in Eq.~\eqref{eq:metric}. Invariance of the mode phase then gives the dimensionless-frequency conversion in Eq.~\eqref{eq:frequencyconversion}. Applying this conversion to the Weyl-chart geodesic results gives
\begin{equation}
\frac{\delta\Omega_\phi}{\Omega_\phi}=-5A,
\qquad
\frac{\delta\Omega_\theta}{\Omega_\theta}=
\frac{\delta\lambda_L}{\lambda_L}=10A.
\label{eq:dictionary}
\end{equation}

Equivalently, and as a direct check, extremising $b^2(r)$ in the
Regge-Wheeler chart of Eq.~\eqref{eq:RWgauge} gives
\begin{equation}
r_c=\frac32\rs(1-5A),
\label{eq:comparison}
\end{equation}
and reproduces Eq.~\eqref{eq:dictionary} without any additional normalisation bookkeeping. These results agree with Ref.~\cite{Cardoso:2021qqu} after identifying its tidal parameter as
$\epsilon_{\rm CF}=-A$.

The wave result also admits a general fixed-$\mu$ eikonal interpretation and the finite-$\ell$ angular coefficient $c_{\ell m}$ can be related to the eikonal orbit-averaged tide. Holding $\mu=m/\ell$ fixed as $\ell\to\infty$, the angular coefficient becomes  
\begin{equation}
\clm=\frac{1-3\mu^2}{4}+ {\cal O}(\ell^{-1})= P_2(0)P_2(\mu) + {\cal O}(\ell^{-1}).
\label{eq:clmeik}
\end{equation} 
This is equivalently the orbit average of the quadrupolar tide along the
corresponding unperturbed great circle. Defining the effective eikonal rates
through
\begin{equation}
\omega_R \sim \left(\ell+\frac12\right)\Omega_c(\mu),
\quad
-{\rm Im}\,\omega \sim \left(n+\frac12\right)\lambda_L(\mu),
\end{equation}
Eqs.~\eqref{eq:master} and~\eqref{eq:kappaeik} give
\begin{equation}
\frac{\delta\Omega_c}{\Omega_c}=\frac52A(1-3\mu^2),
\qquad
\frac{\delta\lambda_L}{\lambda_L}=-5A(1-3\mu^2).
\label{eq:inclined}
\end{equation}
In geometric optics, $\mu\simeq L_z/L=\cos\tilde\theta$, where $\tilde\theta$ is the
inclination of the orbital angular momentum relative to the tidal symmetry
axis. Thus $|\mu|=1$ corresponds to an equatorial orbit where 
$\Omega_c=\Omega_\phi$, so Eq.~\eqref{eq:inclined} reduces precisely to
Eq.~\eqref{eq:dictionary}. In addition, the ratio of the transverse
instability scale to the azimuthal frequency is independent of the
normalisation as shown in Eq.~\eqref{eq:ratesplit}.

The same agreement appears directly in the near-equatorial spectrum. For
fixed $j=\ell-|m|={\cal O}(1),$ the exact angular coefficient has the expansion
\begin{equation}
\clm\big|_{|m|=\ell-j}=-\frac12+\frac{3(j+\frac12)}{2\ell}+{\cal O}(\ell^{-2}),
\end{equation}
and hence
\begin{equation}
\clm\left(\ell+\frac12\right)=-\frac{|m|}{2}+\left(j+\frac12\right)+{\cal O}(\ell^{-1}).
\label{eq:clmneareq}
\end{equation}

Restoring the horizon radius in Eq.~\eqref{eq:kappaeik} and using
$\Omega_0=2/(3\sqrt3\,\rs)$, the wave calculation gives
\begin{align}
\delta\omega
={}&
A\Omega_0
\left[-5|m|+10\left(j+\frac12\right)-i\,5(2n+1)\right]
\nonumber\\
&+{\cal O}(A\Omega_0\ell^{-1},A^2\Omega_0).
\label{eq:neareq}
\end{align}
Equivalently,
\begin{align}\nonumber
\delta\omega=&|m|\,\delta\Omega_\phi
+\left(j+\frac12\right)\delta\Omega_\theta
-i\left(n+\frac12\right)\delta\lambda_L
\\
&+{\cal O}(A\Omega_0\ell^{-1},A^2\Omega_0),
\end{align}
using Eq.~\eqref{eq:dictionary}. 
The equality of the polar level spacing and the radial resonance spacing is
the wave counterpart of the tracelessness condition~\eqref{eq:trace}.  Correspondingly, the same coefficient $20\sqrt3/9$ controls the real transverse and imaginary radial level
spacings in Eq.~\eqref{eq:kappaeik}, reproducing $\Omega_\theta=\lambda_L$ through ${\cal O}(A)$.

Chart-dependent coordinate locations differ, as they must. In particular, the
photon ring lies at $r_{c,\rm Weyl}=\frac32r_{s,\rm Weyl}(1-4A)$ 
in the original Weyl chart, whereas in the Regge-Wheeler chart of
Eq.~\eqref{eq:RWgauge} it lies at $r_{c,\rm RW}=\frac32r_{s,\rm RW}(1-5A).$
Here $r_{s,\rm Weyl}$ is the horizon parameter appearing in the exact metric,
while $r_{s,\rm RW}$ is the physical, renormalised horizon radius.  Their
difference reflects both the ${\cal O}(A)$ radial coordinate transformation
and the horizon-radius redefinition derived in Appendix \ref{App:A1}. The properly normalised orbital frequencies and instability rates nevertheless agree.

The two approaches are complementary. Within the exact Weyl tidal
geometry, the geodesic and Penrose-limit analyses can be carried out
nonperturbatively in $A$ and can therefore capture genuinely nonlinear
structures. 
The wave calculation is perturbative in $A$, but covers the finite-$\ell$ multipoles analysed explicitly here, all their $m$ components, and both parity sectors through ${\cal O}(A)$. In
the eikonal limit, it identifies the $(1-3\mu^2)$ dependence as the continuum
limit of the Wigner-Eckart coefficient $c_{\ell m}$.

By an independent string probe construction an effective temperature $T_{\rm ring}$ is associated with the photon-ring region as $\lambda_L=2\pi T_{\rm ring}$ \cite{Giataganas:2026ctn,Giataganas:2026fop}. The resulting relation  has the same algebraic form as saturation of the chaos bound \cite{Maldacena:2015waa}, and offers a thermal interpretation of the eikonal spectrum~\cite{Giataganas:2026ctn,Giataganas:2026fop}. 
Independently of this analogy, the wave calculation contains the same
instability scale: the least-damped eikonal mode satisfies
$-{\rm Im}\,\omega=\lambda_L/2.$ For a physical companion, $A<0$, the equatorial azimuthal frequency
increases while $\lambda_L$ decreases. The equatorial photon ring therefore
rotates faster but is less unstable, and the corresponding eikonal modes
decay more slowly.

\section{Conclusions}
\label{section:8}
We have derived the diagonal odd- and even-parity gravitational perturbation
equations of a Schwarzschild black hole immersed in a static quadrupolar
vacuum tide and computed the first order shifts of the associated
analytically continued local resonances in closed contour-integral form.
Each shift factorises into one reduced complex coefficient for each
$(n,\ell)$, multiplied by the universal quadrupolar Zeeman pattern
$c_{\ell m}$. For the $\ell=2$ and $\ell=3$ sectors analysed explicitly,
the reduced coefficient is identical in the two parity channels. Moreover, 
$\sum_{m=-\ell}^{\ell}\delta\omega_{n\ell m}=0,$ 
so the tide splits each azimuthal multiplet without shifting its unweighted
centroid at first order. For the dominant $(n,\ell,m)=(0,2,2)$ mode, the
fractional shifts are given in Eq.~\eqref{eq:fractional}.

A pure vacuum linear tide therefore acts as a traceless and, in the sectors
examined, parity-blind spectral splitting field. Matter environments, by
contrast, can shift the centroid and break the Schwarzschild parity
degeneracy~\cite{Leung:1997was,Li:2023ulk}. The same happens for higher-order tides, since at second order in the tidal amplitude, feedback from off-diagonal sidebands can in principle shift the centroid and lift the parity degeneracy.

The eikonal regime provides an independent geometric interpretation of the
wave result. After applying the horizon-radius and Killing-time
normalisation dictionary, the WKB shifts agree with the orbital and
transverse rates obtained from the Penrose limit. Ricci-flatness makes the
two-dimensional transverse curvature matrix traceless and enforces
$\Omega_\theta=\lambda_L$, while the azimuthal orbital frequency
$\Omega_\phi$ shifts independently. The wave coefficient
$\kappa^{\rm eik}_{n\ell}$ reproduces separately the azimuthal contribution,
the stable polar level spacing, and the unstable radial resonance spacing.
Its spin independence at leading eikonal order reflects the universal
propagation of massless fields along null rays, while
$c_{\ell m}\to(1-3\mu^2)/4$ gives the orbit-averaged quadrupolar tide. In
particular, the least-damped eikonal mode satisfies 
$-\operatorname{Im}\omega
=\lambda_L/2
+{\cal O}(\ell^{-1}).$
For a physical companion, $A<0$, the equatorial photon ring rotates faster
but is less unstable, and the corresponding eikonal modes decay more slowly.

The deformed master equations, the intertwining construction associated with
Eq.~\eqref{eq:intertwiner}, and the rational Sturm--Liouville weight
\eqref{eq:weight} form the analytic backbone of these results and provide a
starting point for several natural extensions. The off-diagonal
same-sector $\ell\pm2$ and polar--axial $\ell\pm1$ sidebands already enter
the eigenfunctions at ${\cal O}(A)$ and feed back into the spectrum at
${\cal O}(A^2)$. Other extensions include dynamical and non-axisymmetric
tides, rotating black holes, and global matching of the local tidal geometry
to a time-dependent binary spacetime. Such a matching is necessary to turn
the static resonance shifts derived here into complete observable ringdown
waveforms, including their excitation amplitudes.

These results may also have phenomenological implications for the
identification of subsolar compact objects. Subsolar compact binaries are
often discussed as potential evidence for primordial black
holes~\cite{Riotto:2024ayo}, while material stars and other compact objects
can occupy the same mass range and exhibit finite-size tidal effects. In
four-dimensional general relativity, vacuum black holes instead have a
vanishing intrinsic static Love response, including nonlinearly for
Schwarzschild and Kerr black holes
~\cite{Kehagias:2024rtz,Gounis:2024hcm}. Environmental disturbances can
nevertheless generate an effective tidal response and complicate this
distinction~\cite{DeLuca:2024uju}.

The spectral effect computed here is complementary to the Love response:
the Love number measures the multipole induced on the compact object,
whereas the resonance shift measures how the applied ambient curvature
modifies its dynamical spectrum. In systems where a quasi-static external
tide persists during the observable ringdown, a joint analysis of inspiral
tidal deformability and environment-induced ringdown splitting could help
separate the intrinsic response of the compact object from the external
geometry in which it resides. The central conclusion is that even
the cleanest astrophysically relevant environment, a pure vacuum tide, need not leave the
ringdown spectrum unchanged: it produces a calculable, traceless azimuthal
splitting while preserving the first order parity degeneracy in the sectors
studied here.

\vspace{0.9em}
\centerline{\bf Acknowledgments}\vspace{0.9em}
\noindent
D.G.  acknowledges support from  the National Science and Technology Council (NSTC) of Taiwan with the Young Scholar Columbus Fellowship grant 114-2636-M-110-004 and 115-2112-M-110-010. 
A.R. acknowledges support from the Swiss National
Science Foundation (project number CRSII5\_213497) and from the Boninchi
Foundation for the project ``PBHs in the Era of GW Astronomy''.

\setcounter{equation}{0}

\begin{center}
    \Large\bfseries Appendices
\end{center}
\addcontentsline{toc}{section}{Appendices}
\appendix
\section{Linearised background, angular factor, and radial conventions in Regge-Wheeler gauge}
\label{App:A1}

Expanding the exact metric~\eqref{eq:metric} to first order in $A$ and temporarily setting $r_{s,\rm Weyl}=1$, the quadrupole sector reads, in the original
Weyl chart, with ${\cal W}=4r^{2}-4r+\tfrac23$ and $y=2r-1$,
\begin{align}\nonumber
\delta g_{tt}=&-2Af\,{\cal W}P_{2}\,,\qquad
\delta g_{rr}=-\frac{2A}{f}\Big({\cal W}-\tfrac43 y\Big)P_{2}\,,\\
\delta g_{\theta\theta}=&-2Ar^{2}\Big({\cal W}-\tfrac43y\Big)P_{2}\,,\qquad
\delta g_{\phi\phi}=-2Ar^{2}\sin^{2}\theta\,{\cal W}P_{2}\,,
\label{eq:weylraw}
\end{align}
together with a monopole piece in the $(rr,\theta\theta)$ components.
Using the even-parity gauge vector $\xi_\mu$ with components \cite{Kehagias:2024rtz}
\begin{equation}
\xi_{r}=\tfrac43\,A\,r^{2}\,P_{2}\,,\qquad
\xi_{\theta}=\tfrac23\,A\,r^{2}(1-2r)\,\partial_{\theta}P_{2}\,,
\label{eq:gaugevec}
\end{equation}
the transformation $\delta g_{\mu\nu}\rightarrow
\delta g_{\mu\nu}-2\nabla_{(\mu}\xi_{\nu)}$ removes the
trace-free angular amplitude and brings the quadrupolar perturbation to
Regge-Wheeler gauge.

After this transformation, the perturbation is organised in the multipole
decomposition
\begin{equation}
\delta g_{tt}=f\big[H_{0}^{(0)}+H_{0}^{(2)}P_{2}\big]\,,\qquad
\delta g_{rr}=f^{-1}\big[H_{2}^{(0)}+H_{2}^{(2)}P_{2}\big]\,,\qquad
\delta g_{AB}=r^{2}\big[K^{(0)}+K^{(2)}P_{2}\big]\gamma_{AB}\,,
\label{eq:multidec}
\end{equation}
where $A,B\in\{\theta,\phi\}$ and $\gamma_{AB}$ is the unit-sphere metric.
The quadrupolar Regge-Wheeler-gauge functions quoted in
Eq.~\eqref{eq:RWgauge} of the main text are
\begin{equation}
H_{0}^{(2)}=H_{2}^{(2)}=\eps\,r^{2}f\,,\qquad
K^{(2)}=\eps\big(r^{2}-\tfrac12\big)\,,\qquad \eps=-8A\,,
\label{eq:H0H2K}
\end{equation}
which agrees with the linearisation of the exact solution of
Ref.~\cite{Kehagias:2024rtz}. Up to the overall tidal amplitude, the
monopole mass shift and trivial residual gauge transformations, it is
the unique growing static even-parity $\ell=2$ vacuum solution regular
at the horizon. Restoring the Weyl radius at this order, the angular amplitude has radial
dependence $K^{(2)}\propto r^2-r_{s,\rm Weyl}^2/2$. After the normalisation below this is equivalently
$K^{(2)}\propto r^2-2M^2$ through ${\cal O}(A)$, the horizon-regular tidal
profile of Ref.~\cite{Binnington:2009bb}.

The monopole sector is treated separately. In the decomposition
\eqref{eq:multidec}, the Weyl chart gives
\begin{equation}
H_{0}^{(0)}=0\,,\qquad
H_{2}^{(0)}=-\tfrac83\,A\,y\,,\qquad
K^{(0)}=-\tfrac43\,A\,y\,,
\label{eq:monoampl}
\end{equation}
where the $tt$ component carries no monopole because $\mathcal U$ multiplies
$P_{2}$ only, whereas $k_{0}\propto\sin^{2}\theta$ contains an $\ell=0$
part and enters the $(rr,\theta\theta)$ components but not
$g_{\phi\phi}$. Because $H_{0}^{(0)}$ vanishes while $H_{2}^{(0)}$ does
not, this sector cannot be removed by a radial gauge transformation
together with a mass shift alone, for which
$H_{0}^{(0)}=H_{2}^{(0)}=\delta\rs/(r-1)$: the $K^{(0)}$ equation fixes
the monopole gauge vector
\begin{equation}
\xi_{r}^{(0)}=-\frac{2A}{3}\,\frac{r^{2}(2r-1)}{r-1}
\label{eq:monogauge}
\end{equation}
with no freedom left, after which the $rr$ equation gives the constant
$\delta\rs=-\tfrac23A$, whereas the $tt$ equation would require the
$r$-dependent value $-\tfrac23A\,(2r-1)$. The obstruction is removed by
the residual normalisation freedom of the Killing time, which is not
fixed by any asymptotic condition because the tidal spacetime is not
asymptotically flat. With Eq.~\eqref{eq:normalisation}  
all components match and Eq.~\eqref{eq:RWgauge} follows. Two invariants
confirm the mass shift: the horizon area of Eq.~\eqref{eq:metric} is
${\cal A}_H=4\pi r_{s,\rm Weyl}^2(1-\tfrac43A)$, 
so the areal horizon radius is $r_{s,\rm RW}=r_{s,\rm Weyl}(1-\tfrac23A)$;
and the Misner--Sharp mass of the monopole sector is  $2M_{\rm MS} =r_{s,\rm Weyl}\left(1-\frac{2A}{3}\right)+\mathcal O(A^2)=r_{s,\rm RW}+\mathcal O(A^2)$, independent of $r$.
From this point onward, $\rs$ denotes this physical, renormalised horizon radius and $t$ the correspondingly normalised Killing time, and we again set $\rs=1$. Since this parameter shift is already $\mathcal O(A)$, it does not modify the quadrupolar amplitudes of Eq.~\eqref{eq:H0H2K} at the order considered.

\subsection{The angular factor $\clm$} 
The tidal background is axisymmetric and transforms as the axisymmetric $L=2$ component $Y_{20}$ 
of a spherical tensor. Consequently, $m$ is conserved, and the diagonal $\mathcal O(A)$ contribution obtained by projecting the perturbation equations back onto the same $(\ell,m)$
sector is a diagonal matrix element of a rank-two operator. The
Wigner-Eckart theorem gives
\begin{equation}
\bigl\langle \ell m\big|\mathcal T^{(2)}_{0,a}\big|\ell m\bigr\rangle=(-1)^{\ell-m}
\begin{pmatrix}
\ell & 2 & \ell\\
-m & 0 & m
\end{pmatrix}
\bigl\langle \ell\big\|
\mathcal T^{(2)}_{a}
\big\|\ell\bigr\rangle ,
\label{eq:wed}
\end{equation}
where $a$ labels the projected Einstein equation and the parity sector.
The reduced matrix element in Eq.~\eqref{eq:wed} may depend on $a$ and  $\ell$, but it is independent of $m$. Thus the Wigner-Eckart theorem fixes the common $m$-dependence of the complete diagonal projected operators. It does not imply that the individual scalar, vector, and tensor angular integrals are equal.

We normalise this universal $m$-dependence using the diagonal matrix
element of the scalar multiplication operator $P_2(\cos\theta)$,
\begin{align}
\clm&=\int\dd\Omega\,\overline{Y}_{\ell m}\,P_2(\cos\theta)\,
Y_{\ell m}=\sqrt{\frac{4\pi}{5}}\int\dd\Omega\,\overline{Y}_{\ell m}\,Y_{20}\,
Y_{\ell m}\nonumber\\
&=
(-1)^m(2\ell+1)
\begin{pmatrix}
\ell & 2 & \ell\\
0 & 0 & 0
\end{pmatrix}
\begin{pmatrix}
\ell & 2 & \ell\\
-m & 0 & m
\end{pmatrix}.
\label{eq:clm3j}
\end{align}
All remaining $\ell$-dependent reduced coefficients are included in
the corresponding radial operators. The projections with
$\ell'\neq\ell$ determine the sidebands induced in the perturbed
eigenfunction. In the absence of accidental degeneracies, these
off-diagonal components do not contribute to the first order
frequency shift.
Finally, using
\begin{equation}
\int\dd\Omega\,
|Y_{\ell m}|^2\cos^2\theta=\frac{1}{3}+\frac{2}{3}
\frac{\ell(\ell+1)-3m^2}{(2\ell-1)(2\ell+3)}
\end{equation}
we find that the angular factor $\clm$ is calculated to be
\begin{equation}
\clm=\frac{\ell(\ell+1)-3m^2}{(2\ell-1)(2\ell+3)}.
\label{eq:clmclosed}
\end{equation}

Three consequences follow directly from Eq.~\eqref{eq:clmclosed}. First,
the angular pattern has vanishing trace over the $(2\ell+1)$ values of $m$,
\begin{align}
\sum_{m=-\ell}^{\ell}\clm&=\frac{(2\ell+1)\ell(\ell+1)-3\sum_{m=-\ell}^{\ell}m^2}
{(2\ell-1)(2\ell+3)}=0,
\label{eq:clmtrace}
\end{align}
where
$\sum_{m=-\ell}^{\ell}m^2=\ell(\ell+1)(2\ell+1)/3$.
Consequently, for a first order shift of the form
$\delta\omega_{n\ell m}=A\clm\kappa_{n\ell}$,
\begin{equation}
\frac{1}{2\ell+1}\sum_{m=-\ell}^{\ell}\delta\omega_{n\ell m}=0,
\end{equation}
so the centroid of the multiplet is unchanged at $\mathcal O(A)$.

Second, $c_{\ell,-m}=c_{\ell m}$, and hence the modes with $m$ and $-m$
receive the same first order shift. Third, all the $m$-dependence is fixed
by the diagonal matrix element of the $M=0$ component of the rank-two
tidal tensor. The radial dynamics determines only the reduced coefficient
$\kappa_{n\ell}$ and therefore cannot modify the relative splitting between
the different values of $m$.
For $\ell=2$ one finds
\begin{equation}
c_{2,0}=\frac{2}{7},
\qquad
c_{2,\pm1}=\frac{1}{7},
\qquad
c_{2,\pm2}=-\frac{2}{7},
\end{equation}
which indeed satisfies
$c_{2,0}+2c_{2,1}+2c_{2,2}=0$, whereas
for $\ell=3$,
\begin{equation}
c_{3,0}=\frac{4}{15},
\qquad
c_{3,\pm1}=\frac{3}{15},
\qquad
c_{3,\pm2}=0,
\qquad
c_{3,\pm3}=-\frac{5}{15}.
\end{equation}
These are the relative splittings displayed in Fig.~\ref{plot}. In particular,
$c_{3,\pm2}=0$, so the $|m|=2$ modes receive no diagonal first order
frequency shift.

\subsection{The radial mode functions}
For the multipoles $\ell=2,3$ considered below, projection onto the
diagonal $\ell'=\ell$ sector and elimination of the constraint variables leave one propagating radial degree of freedom in each parity sector. This concerns the diagonal problem that determines the first order frequency shift, whereas the full perturbed metric also contains the off-diagonal sidebands sourced at
$\mathcal O(\eps)$. In the odd sector we use the Regge-Wheeler-type master function
\begin{equation}
\psi(r)=\frac{f(r)}{r}h_1(r),
\label{eq:psidef}
\end{equation}
where $h_1$ is the axial metric amplitude of
Eq.~\eqref{eq:axialansatz}. Its diagonal equation can be written
schematically as
\begin{equation}
\bigl(f\psi'\bigr)'+\frac{\omega^2-V^-_\ell}{f}\,\psi+\eps\clm\,{\cal D}^-_\ell(r,\omega,\partial_r)\psi=0,
\label{eq:oddmasterschematic}
\end{equation}
where ${\cal D}^-_\ell$ is the first order tidal differential operator.
For $\ell=2$ this is Eq.~\eqref{eq:oddmaster} of the main text, and  the
corresponding $\ell=3$ operator is given below.

The even sector is described by the metric amplitude $K$, which obeys its own second order equation after the remaining even-parity amplitudes have been eliminated. It should not be identified with $\psi$, but instead, the odd and even solutions are related, through $\mathcal O(\eps)$, by the deformed intertwining map, which will be derived in Appendix \ref{App:A3},
\begin{equation}
K=\bigl(X_1+\eps\clm Y_1\bigr)\psi+\bigl(X_2+\eps\clm Y_2\bigr)\psi'.
\label{eq:deformedmapsummary}
\end{equation}
At $\eps=0$, this reduces to the usual composition of the
Chandrasekhar--Detweiler transformation with the reconstruction of the
even-parity metric perturbation. In the contour perturbation theory, the
odd-sector integrals are evaluated on $\psi^{(0)}_{n\ell}$, whereas the
even-sector integrals are evaluated on
\begin{equation*}
K^{(0)}_{n\ell}=X_1\bigl(r,\omega^{(0)}_{n\ell}\bigr)\psi^{(0)}_{n\ell}+X_2\bigl(r,\omega^{(0)}_{n\ell}\bigr){\psi^{(0)}_{n\ell}}',
\end{equation*}
obtained from the undeformed intertwining map.

The contour integrals used to calculate the first order frequency shift are
evaluated on the unperturbed Schwarzschild quasinormal mode
$\psi^{(0)}_{n\ell}$ at $\omega=\omega^{(0)}_{n\ell}$, not on a solution of
the tidally deformed equation, and with time dependence $e^{-i\omega t}$, its
Leaver representation in units $\rs=1$ is
\begin{equation}
\psi^{(0)}_{n\ell}(r)=e^{i\omega r}(r-1)^{-i\omega}r^{2i\omega}\sum_{k=0}^{\infty}a_k\left(\frac{r-1}{r}\right)^k,
\qquad
\omega=\omega^{(0)}_{n\ell}.
\label{eq:leaversummary}
\end{equation}
It behaves as $e^{-i\omega r_*}$ at the horizon and as $e^{+i\omega r_*}$ at infinity, corresponding respectively to ingoing and outgoing boundary conditions. In the contour calculation these conditions
are imposed by analytic continuation to the complex radial contour.

The functions denoted by $\psi_r(X_r)$ and $\psi_\perp(X_\perp)$ in the Penrose-limit analysis have a different meaning. They are the radial and transverse separation factors of the
local plane-wave field in Eqs.~\eqref{eq:normal} and~\eqref{eq:inverted}. The former is an inverted oscillator labelled by the overtone number $n$, while the latter is a stable oscillator labelled
by $j$. They are not Regge-Wheeler or Zerilli master functions and are related to the black hole perturbation only in the leading geometric-optics limit.

\section{Odd parity: projection and reduction}
\label{App:A2}
The tidally deformed background is $g^{\rm bg}_{\mu\nu}=g^{\rm Schw}_{\mu\nu}+\eps h^{\rm tide}_{\mu\nu}+\mathcal O(\eps^2)$, where the physical tidal amplitude is
$\eps=-8A$ and the tidal deformation  $h^{\rm tide}_{\mu\nu}$ is given explicitly by
\begin{equation*}
h^{\rm tide}_{\mu\nu}\dd x^\mu\dd x^\nu=P_2(\cos\theta)
\left[r^2f^2\,\dd t^2+r^2\,\dd r^2+r^2\left(r^2-\frac12\right)
\left(\dd\theta^2+\sin^2\theta\,\dd\phi^2\right)
\right].
\end{equation*}
We introduce now an independent bookkeeping parameter $b$ for the dynamical
perturbation and write
\begin{align}
&R_{\mu\nu}
\big[g_{\rm Schw}+\eps h^{\rm tide}+b h^{\rm dyn}\big]
=\eps\,\delta R_{\mu\nu}[h^{\rm tide}]
+b\,\delta R_{\mu\nu}[h^{\rm dyn}]+\eps b\,\delta^2R_{\mu\nu}[h^{\rm tide},h^{\rm dyn}]
+\mathcal O(\eps^2,b^2).
\label{eq:2param}
\end{align}
The term independent of $\eps$ and $b$ vanishes (Schwarzschild background), whereas the term linear in $\eps$ vanishes because
$h^{\rm tide}_{\mu\nu}$ satisfies the linearised vacuum equations,
$\delta R_{\mu\nu}[h^{\rm tide}]=0$. The equation linear in the dynamical
perturbation is therefore
\begin{equation*}
{\cal E}_{\mu\nu}\equiv\delta R_{\mu\nu}[h^{\rm dyn}]+\eps\,\delta^2R_{\mu\nu}[h^{\rm tide},h^{\rm dyn}]=0.
\end{equation*}

Let $A,B\in\{\theta,\phi\}$ denote indices on the unit two-sphere, whose
metric and volume form are $\gamma_{AB}$ and $\varepsilon_{AB}$, respectively.  The axial vector harmonic associated with $Y_{\ell m}$ is the vector $S^A=\gamma^{AB}S_B$, where
\begin{equation*}
S_A\equiv -\varepsilon_A{}^{B}D_BY_{\ell m},
\end{equation*}
and $D_A$ is the covariant derivative associated with $\gamma_{AB}$.
Using the explicit form of the metric $\gamma_{AB}$ we find that 
\begin{equation*}
S^A\partial_A=-\frac{\partial_\phi Y_{\ell m}}{\sin\theta}\,\partial_\theta
+\frac{\partial_\theta Y_{\ell m}}{\sin\theta}\,\partial_\phi, 
\end{equation*}
and satisfies
$\int_{S^2}\dd\Omega\,\overline S^A S_A=\ell(\ell+1)$.
In Regge-Wheeler gauge, the odd-parity dynamical perturbation is
\begin{equation}
h^{\rm dyn}_{\mu\nu}\dd x^\mu\dd x^\nu=2e^{-i\omega t}\big[h_0(r)\,\dd t+h_1(r)\,\dd r\big]S_A\dd x^A,
\label{eq:axialansatz}
\end{equation}
and defining the axial tensor harmonic by
\begin{equation*}
S_{AB}\equiv D_{\!(A}S_{B)}, \qquad S^{AB}\equiv\gamma^{AC}\gamma^{BD}S_{CD},
\end{equation*}
the three axial projections of ${\cal E}_{\mu\nu}=0$ are
\begin{equation}
E^{(t)}=\int_{S^2}\dd\Omega\,\overline S^{A}{\cal E}_{tA}, \qquad
E^{(r)}=\int_{S^2}\dd\Omega\,\overline S^{A}{\cal E}_{rA}, \qquad 
E^{(T)}=\int_{S^2}\dd\Omega\,\overline S^{AB}{\cal E}_{AB}.
\label{eq:axproj}
\end{equation}
At $\eps=0$, the equations $E^{(T)}=0$ and $E^{(r)}=0$ give,
respectively, for $\ell=2$,
\begin{equation}
i\omega h_0+f(fh_1)'=0,
\qquad
i\omega\left(h_0'-\frac{2h_0}{r}\right)-\left(\omega^2-\frac{4f}{r^2}\right)h_1=0.
\label{eq:RWsystem}
\end{equation}
The remaining projection $E^{(t)}=0$ is their differential consequence.

At first order, the mixed curvature contains products of the quadrupolar
background and the axial harmonics. After retaining the diagonal
$\ell'=\ell$ contribution, the angular integrals factorise into
$c_{\ell m}$. With the same radial normalisation,
$E^{(T)}=0$ and $E^{(r)}=0$ become, respectively,
\begin{align}\nonumber
&i\omega\,h_{0}+f\,(fh_{1})'
+\eps\,c_{2m}\Big[-\frac{i\omega(10r^{2}-4r-3)}{4}\,h_{0}
-\frac{(r-1)^{2}(2r^{2}+4r-3)}{4r^{2}}\,h_{1}'\\
&-\frac{(r-1)(4r^{3}+2r^{2}-3)}{4r^{3}}\,h_{1}\Big]=0\,,
\label{eq:defT}\\[3pt]
&i\omega\Big(h_{0}'-\frac{2}{r}h_{0}\Big)-\Big(\omega^{2}-\frac{4f}{r^{2}}\Big)h_{1}
+\eps\,c_{2m}\Big[\frac{i\omega r(r-1)}{2}\,h_{0}'-i\omega(2r-1)\,h_{0}\nonumber\\
&
-\frac{(r-1)(\omega^{2}r^{4}+4r^{2}+12r-8)}{2r^{3}}\,h_{1}\Big]=0\,.
\label{eq:defR}
\end{align}
Equation~\eqref{eq:defT} remains algebraic in $h_{0}$, and  its perturbative solution is
\begin{equation}
h_{0}=\frac{i}{\omega}\,f\,(fh_{1})'+\eps c_{2m}
\frac{i(r-1)^{3}\big(2r\,h_{1}'-h_{1}\big)}{\omega r^{2}}.
\label{eq:h0elim}
\end{equation}
Substituting Eq.~\eqref{eq:h0elim} into the dynamical equation
\eqref{eq:defR} and introducing the Regge-Wheeler-type master variable as in \eqref{eq:psidef} 
we first normalise the $\eps^{0}$ part to the standard self-adjoint
Regge-Wheeler form. Any occurrence of $\psi''$ inside the term already
proportional to $\eps$ may then be eliminated using the unperturbed
Regge-Wheeler equation. This on-shell reduction changes the first order
operator only by a term of the form ${\cal Q}\hat L_{\ell}^{(0)}$, where
$\hat L_{\ell}^{(0)}\psi^{(0)}=0$. It therefore makes no contribution to
the numerator of the first order shift formula~\eqref{eq:LSG} and leaves
the frequency shift unchanged. The resulting equation is the deformed
master equation~\eqref{eq:oddmaster}.

The following three independent facts control this reduction. (i)~The remaining
projection $E^{(t)}=0$ is not an independent field equation. The
linearised contracted Bianchi identity relates it to $E^{(T)}=0$ and
$E^{(r)}=0$. After substituting Eq.~\eqref{eq:h0elim} and the master
equation, $E^{(t)}$ vanishes identically at both $\eps^{0}$ and
$\eps^{1}$, as verified in exact rational arithmetic. (ii)~The projections were evaluated independently for
$m=0,1,2$ using explicit spherical harmonics. Defining the complete
diagonal master operator by
\begin{equation*}
\hat L_{\ell m}=\hat L_{\ell}^{(0)}+\eps\,\hat F_{1}^{(\ell m)},
\end{equation*}
the three first order operators satisfy
\begin{equation}
\hat F_{1}^{(\ell m)}=\frac{\clm}{c_{\ell0}}\,\hat F_{1}^{(\ell0)}.
\label{eq:WEop}
\end{equation}
This is the Wigner-Eckart factorisation of the complete projected
operator. It does not require the individual scalar, vector and tensor
angular integrals entering the calculation to be identical.
(iii)~The part of the $\ell=2$ tidal operator proportional to
$\omega^{2}$ determines the correction to the outgoing characteristic
phase. Writing a local outgoing solution as
$\psi_{\rm out}\sim\exp[\int^{r}k(s)\dd s]$, the characteristic root
through first order in $\eps$ is
\begin{equation}
k(r)\simeq i\omega\,\frac{\dd x}{\dd r},
\qquad \frac{\dd x}{\dd r}=\frac{1}{f}\left(1-\eps c_{2m}r^{2}f\right)+\mathcal O(\eps^{2}),
\qquad
x=r_{*}-\eps c_{2m}\frac{r^{3}}{3}+\mathcal O(\eps^{2}),
\label{eq:tortSM}
\end{equation}
which reproduces Eq.~\eqref{eq:tortoise}. 
Because the tidal perturbation grows as $\eps r^{2}$, this expression is
a perturbative large-$r$ phase valid only in the  region $ 1\ll |r|\ll |\eps c_{2m}|^{-1/2}.$  Thus, this large-$r$ phase is valid only in the perturbative overlap region \eqref{eq:overlap}. It is used in Appendix \ref{App:A4} to select the local outgoing direction in the
finite-$\eps$ direct integration of the linearly truncated equations.
It should not be interpreted as a boundary condition imposed at the
non-asymptotically-flat infinity of the exact tidally distorted
spacetime.

Repeating the same projection and reduction for $\ell=3$ gives
\begin{equation}
\big(f\psi'\big)'+\frac{\omega^{2}-V^{-}_{\ell=3}}{f}\,\psi+\eps c_{3m}\left[-3(r-1)\psi'+\frac{3(9r^{2}-5)}{2r^{2}}\,\psi\right]=0,
\label{eq:oddl3}
\end{equation}
where the potential $V_\ell^-$ reads:
\begin{equation}
 V_\ell^-=f\left[\frac{\ell(\ell+1)}{r^2}-\frac{3}{r^3}\right].
\end{equation}
Equation~\eqref{eq:WEop} is again verified independently for
$m=0,\ldots,3$. In particular, because $c_{32}=0$, the complete diagonal
first order deformation vanishes for the $|m|=2$ modes.

Unlike the $\ell=2$ equation, the tidal operator in Eq.~\eqref{eq:oddl3} contains no term proportional to $\omega^{2}$ and therefore generates no $\mathcal O(\eps r^{3})$ correction to the
outgoing characteristic phase. Thus the dephasing in Eq.~\eqref{eq:tortSM} is specific to the diagonal coupling of the quadrupolar tide to the dynamical $\ell=2$ multipole. The invariant statement is the presence or absence of this additional outgoing phase, not the value of an individual coefficient in a chosen master equation. Regular rational redefinitions of the master variable that preserve the boundary class may redistribute coefficients in the radial equation but cannot create or remove this exponential phase.

\section{Even parity: reduction, intertwiner and weight}
\label{App:A3}

The even-parity ansatz in Regge-Wheeler gauge is
\begin{equation}
h^{\rm dyn}_{\mu\nu}\dd x^{\mu}\dd x^{\nu}
=e^{-i\omega t}\,Y_{\ell m}\Big[f\,H_{0}\,\dd t^{2}+2H_{1}\,\dd t\,\dd r+f^{-1}H_{2}\,\dd r^{2}
+r^{2}K\big(\dd\theta^{2}+\sin^{2}\theta\,\dd\phi^{2}\big)\Big].
\label{eq:polaransatz}
\end{equation}
The dynamical Ricci tensor ${\cal E}_{\mu\nu}$ was defined in Appendix \ref{App:A2}, and 
its seven diagonal even-parity projections are
\begin{equation*}
\begin{aligned}
E_{\alpha\beta}&=\int_{S^2}\dd\Omega\,
\overline Y_{\ell m}\,{\cal E}_{\alpha\beta},\qquad 
E_{\alpha V}=\int_{S^2}\dd\Omega\,D^{A}\overline Y_{\ell m}\,{\cal E}_{\alpha A},
\qquad\alpha,\beta\in\{t,r\},
\\
E_{\Omega}&=\int_{S^2}\dd\Omega\,\overline Y_{\ell m}\,\gamma^{AB}{\cal E}_{AB},
\qquad
E_{\rm TF}=\int_{S^2}\dd\Omega\,\overline Z^{AB}{\cal E}_{AB},
\end{aligned}
\end{equation*}
where $D_A$ is the covariant derivative on the unit sphere and
\begin{equation*}
Z_{AB}=\left(D_A D_B+\frac12\ell(\ell+1)\gamma_{AB}\right)Y_{\ell m}
\end{equation*}
is the trace-free polar tensor harmonic. Only the diagonal
$\ell'=\ell$ part of each projection is retained.
At $\mathcal O(\eps^0)$ and for $\ell=2$, the trace-free equation
$E_{\rm TF}=0$ gives $H_0=H_2$, and the equations $E_{tr}=0$,
$E_{tV}=0$ and $E_{rV}=0$ then give, respectively,
\begin{align}
&2\omega r^{2}(r-1)\,K'+\omega r(2r-3)\,K-2\omega r(r-1)\,H_{2}-6i(r-1)\,H_{1}=0\,,
\label{eq:poltr}\\
&i\omega r^{2}\big(H_{2}+K\big)+r(r-1)\,H_{1}'+H_{1}=0\,,
\label{eq:poltV}\\
&i\omega r^{2}\,H_{1}+r(r-1)\big(H_{2}'-K'\big)+H_{2}=0\,,
\label{eq:polrV}
\end{align}
where Eq.~\eqref{eq:polrV} has been written after using $H_0=H_2$. The same reduction can be performed perturbatively through
$\mathcal O(\eps)$. First, the trace-free projection determines $H_0$
in terms of the remaining amplitudes, with
$H_0=H_2+\mathcal O(\eps)$. Second, the tidally corrected versions of
Eqs.~\eqref{eq:poltr}--\eqref{eq:polrV} are solved as a linear system
for $(K',H_1',H_2')$. Third, these expressions, together with the
derivative of the trace-free relation, are substituted into the
angular-trace equation $E_{\Omega}=0$. The result is an algebraic
constraint, namely the tidal deformation of the corresponding
Schwarzschild identity, whose $\mathcal O(\eps^0)$ part is
\begin{equation}
H_{2}=-\frac{4\omega^{2}r^{4}-8r^{2}+6r+3}{2(r-1)(4r+3)}\,K
+\frac{i\,(2\omega^{2}r^{3}-3)}{\omega\,r\,(4r+3)}\,H_{1}\,.
\label{eq:constraint}
\end{equation}
Finally, eliminating $H_2$ with the algebraic constraint leaves the
closed first order system
\begin{equation*}
u'=\bigl(\mathbb M_0+\eps\clm\mathbb M_1\bigr)u,
\qquad
u=
\begin{pmatrix}
K\\ H_1
\end{pmatrix}.
\end{equation*}
The zeroth- and first order matrices $\mathbb M_0$ and $\mathbb M_1$, respectively, are
\begin{equation}
\mathbb M_{0}=
\begin{pmatrix}
-\dfrac{2\omega^{2}r^{4}-3}{r(r-1)(4r+3)} &
\dfrac{2i\,(\omega^{2}r^{3}+6r+3)}{\omega\,r^{2}(4r+3)}\\[10pt]
\dfrac{i\omega r\,(4\omega^{2}r^{4}-16r^{2}+8r+9)}{2(r-1)^{2}(4r+3)} &
\dfrac{2\omega^{2}r^{4}-7r-3}{r(r-1)(4r+3)}
\end{pmatrix},
\label{eq:M0}
\end{equation}
\begin{align}
(\mathbb M_{1})_{KK}&=-\frac{(2r+1)\,\big(4\omega^{2}r^{5}+8\omega^{2}r^{4}+36r^{3}-69r^{2}-6r+42\big)}{2(r-1)(4r+3)^{2}}\,,
\nonumber\\
(\mathbb M_{1})_{KH}&=\frac{i\,\big(16\omega^{2}r^{5}+23\omega^{2}r^{4}+3\omega^{2}r^{3}-3\omega^{2}r^{2}+84r^{3}-3r^{2}-84r-24\big)}{\omega\,r\,(4r+3)^{2}}\,,
\nonumber\\
(\mathbb M_{1})_{HK}&=-\frac{i\omega r}{4(r-1)^{2}(4r+3)^{2}}
\big(32\omega^{2}r^{7}+12\omega^{2}r^{6}-20\omega^{2}r^{5}-12\omega^{2}r^{4}
\nonumber\\
&\hspace{9.2em}-304r^{5}+108r^{4}+360r^{3}-17r^{2}-117r-27\big)\,,
\nonumber\\
(\mathbb M_{1})_{HH}&=\frac{8\omega^{2}r^{6}+20\omega^{2}r^{5}+8\omega^{2}r^{4}+72r^{4}-18r^{3}-39r^{2}+3r-9}{2(r-1)(4r+3)^{2}}\,.
\label{eq:M1}
\end{align}
The repeated factor $4r+3$ is twice the standard Zerilli combination
$\lL r+3M$. Indeed, for $\ell=2$ one has
$\lL=(\ell-1)(\ell+2)/2=2$, while $\rs=2M=1$, and therefore
$4r+3=2\bigl(\lL r+3M\bigr).$
Eliminating $H_{1}$ gives the second order equation $K''+p_{1}K'+p_{0}K=0$ with
\begin{equation}
p_{1}^{(0)}=\frac{\omega^{2}r^{3}+12r^{2}+3r-6}{r(r-1)\,(\omega^{2}r^{3}+6r+3)}\,,\qquad
p_{0}^{(0)}=\frac{\omega^{4}r^{5}+6\omega^{2}r^{3}+6\omega^{2}r^{2}-3\omega^{2}r-36r+36}{(r-1)^{2}\,(\omega^{2}r^{3}+6r+3)}\,,
\label{eq:pzero}
\end{equation}
and first order coefficients, per unit $\eps\,\clm$,
\begin{align}
p_{1}^{(1)}&=-\frac{8\omega^{4}r^{7}+8\omega^{4}r^{6}+96\omega^{2}r^{5}+189\omega^{2}r^{4}+66\omega^{2}r^{3}
-9\omega^{2}r^{2}+252r^{3}+279r^{2}+90r+27}{2\,(\omega^{2}r^{3}+6r+3)^{2}}\,,
\label{eq:p11}\\[4pt]
p_{0}^{(1)}&=-\frac{{\cal N}_{0}(r)}{2\,(r-1)\,(\omega^{2}r^{3}+6r+3)^{2}}\,,
\label{eq:p01}\\
{\cal N}_{0}&=4\omega^{6}r^{9}+30\omega^{4}r^{7}+30\omega^{4}r^{6}+3\omega^{4}r^{5}
-108\omega^{2}r^{5}-99\omega^{2}r^{4}+27\omega^{2}r^{3}
\nonumber\\
&\quad-18\omega^{2}r^{2}+9\omega^{2}r-756r^{3}-108r^{2}+216r+162\,.
\nonumber
\end{align}
The zeros of $\omega^2r^3+6r+3$ are apparent singularities of the
second order equation for $K$. A useful diagnostic is provided by the
partial-fraction residues of $p_1^{(0)}$, which are found to be  $2$ at $r=0$, $1$
at $r=1$ and $-1$ at each of the three roots of
$\omega^2r^3+6r+3$. Their sum vanishes, and hence the integrating
factor that puts the equation in Sturm--Liouville form is the rational
function of Eq. \eqref{eq:weight}. Because $W$ is rational, these apparent roots introduce isolated poles
but no additional branch choices. The contour must avoid the poles,
while the only branch prescription is the one already required by the
quasinormal mode function.

The apparent character of these singularities follows directly from
the rational intertwining map. Substituting
$K=X_1\psi+X_2\psi'$ into the $\mathcal O(\eps^0)$ equation for $K$,
and using
\begin{equation*}
\psi''=-{\cal A}\psi'-{\cal B}\psi,
\qquad
{\cal A}=\frac{f'}{f},
\qquad
{\cal B}=\frac{\omega^2-V^-}{f^2},
\end{equation*}
we may define for convenience the quantities 
${\cal P}$ and ${\cal Q}$ as 
\begin{equation*}
{\cal P}=X_1'-{\cal B}X_2,
\qquad
{\cal Q}=X_1+X_2'-{\cal A}X_2, 
\end{equation*}
so that the coefficients of $\psi$ and $\psi'$ then vanish independently if
\begin{equation}
{\cal P}'-{\cal B}{\cal Q}+p_{1}^{(0)}{\cal P}+p_{0}^{(0)}X_{1}=0\,,\qquad
{\cal P}+{\cal Q}'-{\cal A}{\cal Q}+p_{1}^{(0)}{\cal Q}+p_{0}^{(0)}X_{2}=0\,.
\label{eq:intcond}
\end{equation}
The rational solution relevant here, fixed up to an overall
normalisation, is the map~\eqref{eq:intertwiner}, and  for generic
quasinormal frequencies it maps the ingoing and outgoing
Regge-Wheeler solutions into the corresponding even-parity boundary
classes. At $\eps=0$ it is the composition of the
Chandrasekhar--Detweiler transformation with the standard
reconstruction of the even-parity metric amplitude $K$ from the
Zerilli master function.

One may now ask whether this intertwining relation persists through first
order in the tidal field. To answer this question, we may express $K$ as in \eqref{eq:deformedmapsummary}, where $\psi$ solves the deformed odd-parity master equation, and by defining
$q_1$ and $q_0$ by
\begin{equation*}
{\cal D}^{-}_{\ell}\psi=q_1(r,\omega)\psi'+q_0(r,\omega)\psi,
\end{equation*}
the odd equation in second order normal form has
\begin{equation*}
{\cal A}
\longrightarrow
{\cal A}+\eps\clm\,\frac{q_1}{f},
\qquad
{\cal B}
\longrightarrow
{\cal B}+\eps\clm\,\frac{q_0}{f}.
\end{equation*}
Requiring $K$ to satisfy the even-parity equation with the coefficients
in Eqs.~\eqref{eq:p11}--\eqref{eq:p01}, converts the two intertwining
conditions~\eqref{eq:intcond} into an inhomogeneous system for
$(Y_1,Y_2)$. It admits the following rational solution, unique up to
the common shift
\begin{equation*}
Y_i\longrightarrow Y_i+C(\omega)X_i,
\end{equation*}
which corresponds only to an $\mathcal O(\varepsilon)$ rescaling of $K$. One convenient representative is
\begin{align}
Y_{1}&=-\frac{3\omega^{4}r^{5}+\omega^{4}r^{4}-\omega^{4}r^{3}-24\omega^{2}r^{4}-4\omega^{2}r^{3}
-20\omega^{2}r^{2}-18\omega^{2}r+360r^{2}+90r-90}{32\,\omega^{2}r^{3}}\,,
\label{eq:Y1}\\[3pt]
Y_{2}&=\frac{(r-1)\,\big(10\omega^{2}r^{3}+4\omega^{2}r^{2}-2\omega^{2}r-60r-45\big)}{16\,\omega^{2}r^{2}}. 
\label{eq:Y2}
\end{align}
Substitution of Eqs.~\eqref{eq:Y1}--\eqref{eq:Y2} makes both
intertwining conditions vanish as polynomial identities in $r$ and
$\omega$, as can be directly verified. For $\omega\neq0$, the
deformation introduces no new radial singularities in the exterior
domain. Moreover, $Y_2$ vanishes linearly at the horizon, as does
$X_2$, while $Y_1$ is regular there. At large $r$, the $Y_i$ grow only
polynomially and therefore do not alter the exponential boundary class.
Together with the common characteristic phase~\eqref{eq:tortSM}, this
shows that the map preserves the ingoing and outgoing boundary classes
within the first order tidal problem.
The diagonal even- and odd-parity spectral problems are consequently
intertwined through $\mathcal O(\eps)$, and hence
\begin{equation*}
\kappa^{\rm even}_{n\ell}=\kappa^{\rm odd}_{n\ell}
\end{equation*}
for the multipoles considered here. The agreement of the independent contour evaluations in Appendix \ref{App:A4} is a numerical check of this operator
identity.

At $\eps=0$, the map~\eqref{eq:intertwiner} is the metric-level form
of the Chandrasekhar--Detweiler duality and is ultimately related to
the Teukolsky--Starobinsky identities. There is,
however, no general theorem stating that an arbitrary nonspherical
vacuum deformation must preserve this structure or admit a single
decoupled Teukolsky-type equation~\cite{Teukolsky:1973ha}. In the present problem,
isospectrality follows from the explicit rational solution
\eqref{eq:Y1}--\eqref{eq:Y2} of the deformed intertwining conditions,
rather than from vacuum character alone. The result therefore applies
to the diagonal first order tidal operators derived here. At
$\mathcal O(\eps^2)$, feedback from the off-diagonal sidebands can in
principle lift the degeneracy.

The whole structure persists at $\ell=3$. The reduction gives the apparent factor
$\omega^{2}r^{3}+30r+6$, whose residues in $p_{1}^{(0)}$ are again $(2,1,-1,-1,-1)$, so that
$W=r^{2}(r-1)/(\omega^{2}r^{3}+30r+6)$. The undeformed intertwiner is
\begin{equation}
X_{1}=-\frac{\omega^{2}r^{3}-120r^{2}-6r+6}{200\,r^{3}}\,,\qquad
X_{2}=\frac{(r-1)(10r+3)}{100\,r^{2}}\,,
\label{eq:Xl3}
\end{equation}
with $10r+3=2(\lL r+3M)$ at $\lL=5$, and its tidal deformation again exists in rational form,
\begin{align}\nonumber
Y_{1}=&-\frac{1}{800\,\omega^{2}r^{3}}\big(5\omega^{4}r^{5}+3\omega^{4}r^{4}-4\omega^{4}r^{3}+120\omega^{2}r^{4}-234\omega^{2}r^{3}
-396\omega^{2}r^{2}\\
&+12\omega^{2}r-12\omega^{2}+18000r^{2}+900r-900\big)\,,
\label{eq:Y1l3}\\[3pt]
Y_{2}&=\frac{(r-1)\,\big(25\omega^{2}r^{3}+15\omega^{2}r^{2}+5\omega^{2}r-3\omega^{2}-750r-225\big)}
{200\,\omega^{2}r^{2}}\,,
\label{eq:Y2l3}
\end{align}
with both intertwining conditions vanishing as exact polynomial
identities. Thus the diagonal odd- and even-parity operators are also
isospectral through first order at $\ell=3$.

\section{The shift integrals and their checks}
\label{App:A4}
On the contour ${\cal C}$ the unperturbed mode is represented by the Leaver
series~\cite{Leaver:1985ax}
\begin{equation}
\psi=e^{i\omega r}\,(r-1)^{-i\omega}\,r^{2i\omega}\sum_{k=0}^{\infty}a_{k}\,u^{k}\,,\qquad
u=\frac{r-1}{r}\,,\qquad
\alpha_{k}\,a_{k+1}+\beta_{k}\,a_{k}+\gamma_{k}\,a_{k-1}=0\,,
\label{eq:leaver}
\end{equation}
\begin{align}\nonumber
&\alpha_{k}=(k+1)(k+1-2i\omega)\,,\quad
\beta_{k}=-\big(2k^{2}+2k+1\big)+8i\omega k+4i\omega+8\omega^{2}-\ell(\ell+1)+s^{2}\,,
\\
&\gamma_{k}=(k-2i\omega)^{2}-s^{2}\,.
\label{eq:recur}
\end{align}
 For gravitational perturbations we set $s=2$, and we choose the branch cut
of $(r-1)^{-i\omega}$ along the positive imaginary direction, whereas
the two vertical legs of ${\cal C}$ are taken on opposite sides of this cut. Then, the
minimal Leaver solution converges on both legs and has the analytically
continued outgoing behaviour $\psi\sim e^{i\omega r_*}$~\cite{Lestingi:2026peq}.
To make the decay explicit, we express $\omega$ as
\begin{equation*}
\omega=\omega_R-i\Gamma,\qquad \omega_R>0,\qquad \Gamma>0,
\end{equation*}
so that, up to powers of $r$, the magnitude of the outgoing solution is 
\begin{equation}
|\psi| \sim \exp\!\left[ \Gamma\,{\rm Re}\,r-\omega_R\,{\rm Im}\,r\right]
\qquad
\text{along the vertical legs}.
\label{eq:decay}
\end{equation}
Since ${\rm Re}\,r$ remains bounded on each leg while
${\rm Im}\,r\rightarrow+\infty$, the mode decays exponentially at both
endpoints. These are the analytically continued boundary conditions of the
undeformed Schwarzschild problem. The ingoing horizon behaviour is
encoded in the factor $(r-1)^{-i\omega}$, while the outgoing condition
is imposed by the minimal solution of the recurrence
\eqref{eq:recur}, which gives Leaver's quasinormal-mode condition.

No boundary condition at the non-asymptotically-flat infinity of the exact tidally distorted spacetime enters the first order shift formula, and  the contour calculation instead determines the formal first order coefficient of the analytically continued Schwarzschild resonance. The boundary terms generated by integration by parts vanish because the unperturbed mode, and the first order correction belonging to the same complex-scaled boundary class, decay at the endpoints of ${\cal C}$. ${\cal C}$ descends from $+i\infty$ on one side of the branch cut from $r=1$, encircles the horizon branch point, and 
returns to $+i\infty$ on the opposite side.

The direct finite-$\eps$ checks described below solve equations that
have themselves been truncated at $\mathcal O(\eps)$. At the horizon
the Frobenius exponent remains $-i\omega$ after the mass
renormalisation. At the upper end of the finite numerical contour, the
outgoing direction is selected using the local characteristic root
\begin{equation*}
k(r) \simeq i\omega\,\frac{\dd x}{\dd r},
\end{equation*}
with $x$ given by Eq.~\eqref{eq:tortSM}. Because the tidal expansion
requires $|\eps\clm|\,|r|^2\ll1$, the numerical endpoint must lie below
the turnover scale $|r|\sim|\eps\clm|^{-1/2}.$  Nevertheless,  compatibility with the tidal geometry requires the stronger mode-independent bound of Eq.~\eqref{eq:overlap}.
Stability under changes of this endpoint is included among the
numerical checks. This procedure tests the first order truncated
system, which, however, is not a non-perturbative definition of the spectrum of the
exact non-asymptotically-flat geometry.

The odd-parity coefficient is given by the ratio of quadratures in
Eq.~\eqref{eq:kappaint}. Eliminating $\psi''$ on shell changes the
first order operator by a term proportional to
$\hat L^{(0)}\psi^{(0)}$, which vanishes pointwise. More general
$\mathcal O(\eps)$ field redefinitions and rescalings may also generate
terms in which $\hat L^{(0)}$ acts to the left. Their contour matrix
elements vanish after integration by parts, provided that the
transformation is regular and preserves the endpoint boundary class.
The first order shift is therefore invariant under this restricted
class of master-variable redefinitions.

In the even sector, the unperturbed eigenfunction entering the contour
integrals is
\begin{equation*}
K^{(0)}_{n\ell}= X_1\bigl(r,\omega^{(0)}_{n\ell}\bigr)\psi^{(0)}_{n\ell}+X_2\bigl(r,\omega^{(0)}_{n\ell}\bigr)
{\psi^{(0)}_{n\ell}}'.
\end{equation*}
Writing
$\delta\omega_{n\ell m}=\eps\clm\,\omega^{(1)}_{n\ell}$,
the even-parity shift is
\begin{equation}
\omega^{(1)}_{n\ell}=-\,\frac{\displaystyle \int_{\cal C}\dd r\; W K^{(0)}_{n\ell}
\left[p_1^{(1)}{K^{(0)}_{n\ell}}'+p_0^{(1)}K^{(0)}_{n\ell}\right]}
{\displaystyle
\int_{\cal C}\dd r\;
\left[
-\partial_\omega W\,
 \bigl({K^{(0)}_{n\ell}}'\bigr)^2
+\partial_\omega\bigl(Wp_0^{(0)}\bigr)
 \bigl(K^{(0)}_{n\ell}\bigr)^2
\right]}
\Bigg|_{\omega=\omega^{(0)}_{n\ell}} .
\label{eq:evenshift}
\end{equation}
Here $W$ is given by Eq.~\eqref{eq:weight}, and $p_i^{(1)}$ are the coefficients per unit $\eps\clm$ defined in
Eqs.~\eqref{eq:p11}--\eqref{eq:p01}. Multiplying $W$ by an arbitrary frequency-dependent normalisation does not change the result since the
additional term in $\partial_\omega W$ is proportional to the unperturbed equation and vanishes on shell.
Since $\eps=-8A$, the coefficient used in the main text is
\begin{equation*}
\delta\omega_{n\ell m} = A\clm\,\kappa_{n\ell},
\qquad
\kappa_{n\ell}=-8\,\omega^{(1)}_{n\ell},
\end{equation*}
and the two sectors
give the values already presented in Eqs. \eqref{eq:kappavals} and \eqref{eq:kappal3} as
\begin{align}
\kappa^{\rm odd}_{02}&=\kappa^{\rm even}_{02}=9.98+5.28\,i\,,\qquad
\kappa^{\rm odd}_{12}=\kappa^{\rm even}_{12}=1.85+10.56\,i\,,
\nonumber\\
\kappa^{\rm odd}_{03}&=\kappa^{\rm even}_{03}=13.64+4.45\,i\,.
\label{eq:kappatable}
\end{align}
The quoted values are stable at the $10^{-7}$ level under changes of
the contour, its truncation height and the depth of the Leaver series.
The central odd- and even-parity evaluations coincide to ten internally retained digits. Conservatively, however, their numerical agreement is verified
only to the $10^{-7}$ accuracy established by these stability tests.
The exact equality follows from the intertwining identities of Appendix \ref{App:A3}.
The numerical residual
$
\big\langle K^{(0)}, \hat L^{(0)}(\omega^{(0)})K^{(0)} \big\rangle_{\cal C}
$
is smaller than $10^{-8}$ of the natural scale of the individual
terms, checking the implementation of the weight, contour and
intertwiner.

As a further test, we solve the first order-truncated even-parity system
\begin{equation*}
u'=\bigl(\mathbb M_0+\eta\mathbb M_1\bigr)u, \qquad \eta\equiv\eps\clm,
\end{equation*}
at several small nonzero values of $\eta$. Extrapolation to $\eta=0$
reproduces $\omega_{02}^{(0)}=0.747-0.178\,i$
to $6\times10^{-14}$, while the finite-difference derivative
$\dd\omega/\dd\eta$ agrees with the bilinear-form result to better than
$10^{-7}$. This check is independent of the reduction to the
second order $K$ equation and of the contour perturbation formula,
although it uses the same projected Einstein system and is therefore
not an independent derivation of $\mathbb M_1$ from
$\delta R_{\mu\nu}=0$.

The analogous direct checks in the odd sector agree with the bilinear
prediction at the $10^{-5}$ level for $\ell=2$ and at the
$4\times10^{-7}$ level for $\ell=3$. The remaining discrepancy is
consistent with the quadratic Richardson error of the finite-$\eta$
derivative.
For comparison, the same tide deforms a massless scalar field through
\begin{equation}
\big(r^{2}f\,R'\big)'+\Big(\frac{\omega^{2}r^{2}}{f}-\Lambda\Big)R
+\eps\,\clm\Big[\Big(r^{2}f\big(r-\tfrac12\big)R'\Big)'
+\frac{\omega^{2}r^{2}}{f}\Big(2r^{2}-r-\tfrac12\Big)R\Big]=0\,,\qquad \Lambda=\ell(\ell+1)\,.
\label{eq:scalar}
\end{equation}
Within the same perturbative domain
$|\eps\clm|\,|r^2|\ll1$, its Liouville normal form has the
characteristic coordinate
\begin{equation*}
x= r_* +\eps\clm\,\frac{r^3}{3}+\mathcal O(\eps^2),
\end{equation*}
whose first order correction has the opposite sign to the gravitational
$\ell=2$ result~\eqref{eq:tortSM}.
Writing the normal-form potential as
$\widetilde V=V_0+\eps\clm\,V_1+\mathcal O(\eps^2),$
the coefficient of the first order correction is
\begin{equation}
V_1=-\frac{(r-1)\big(4\Lambda r^2-2\Lambda r-\Lambda-2r^2+2r-2\big)}{2r^3}\,,
\label{eq:V1scalar}
\end{equation}
and the corresponding shifts are
\begin{equation*}
\kappa^{s=0}_{02}=7.20+3.02\,i,
\qquad
\kappa^{s=0}_{03}=11.67+3.42\,i.
\end{equation*}
Thus the spin-two coefficients are not constant rescalings of the
scalar coefficients. Moreover, at $\ell=3$ the gravitational tidal
operator contains no $\omega^2$ correction and hence no
$\mathcal O(\eps r^3)$ modification of the outgoing characteristic
phase, as shown in Eq.~\eqref{eq:oddl3}.

\bibliographystyle{JHEP}
\bibliography{qnm_tidalb}

@article{LIGOScientific:2021sio,
    author = "Abbott, R. and others",
    collaboration = "LIGO Scientific, VIRGO, KAGRA",
    title = "{Tests of General Relativity with GWTC-3}",
    eprint = "2112.06861",
    archivePrefix = "arXiv",
    primaryClass = "gr-qc",
    reportNumber = "LIGO-P2100275",
    doi = "10.1103/PhysRevD.112.084080",
    journal = "Phys. Rev. D",
    volume = "112",
    number = "8",
    pages = "084080",
    year = "2025"
}

@article{Combaluzier-Szteinsznaider:2024sgb,
    author = "Combaluzier-Szteinsznaider, Oscar and Hui, Lam and Santoni, Luca and Solomon, Adam R. and Wong, Sam S. C.",
    title = "{Symmetries of vanishing nonlinear Love numbers of Schwarzschild black holes}",
    eprint = "2410.10952",
    archivePrefix = "arXiv",
    primaryClass = "gr-qc",
    doi = "10.1007/JHEP03(2025)124",
    journal = "JHEP",
    volume = "03",
    pages = "124",
    year = "2025"
}

@article{Kokkotas:1999bd,
    author = "Kokkotas, Kostas D. and Schmidt, Bernd G.",
    title = "{Quasinormal modes of stars and black holes}",
    eprint = "gr-qc/9909058",
    archivePrefix = "arXiv",
    doi = "10.12942/lrr-1999-2",
    journal = "Living Rev. Rel.",
    volume = "2",
    pages = "2",
    year = "1999"
}

@article{Berti:2009kk,
    author = "Berti, Emanuele and Cardoso, Vitor and Starinets, Andrei O.",
    title = "{Quasinormal modes of black holes and black branes}",
    eprint = "0905.2975",
    archivePrefix = "arXiv",
    primaryClass = "gr-qc",
    doi = "10.1088/0264-9381/26/16/163001",
    journal = "Class. Quant. Grav.",
    volume = "26",
    pages = "163001",
    year = "2009"
}

@article{Berti:2025hly,
    author = "Berti, Emanuele and others",
    editor = "Berti, Emanuele and Cardoso, Vitor and Carullo, Gregorio",
    title = "{Black hole spectroscopy: from theory to experiment}",
    eprint = "2505.23895",
    archivePrefix = "arXiv",
    primaryClass = "gr-qc",
    doi = "10.1088/1361-6382/ae59e2",
    journal = "Class. Quant. Grav.",
    volume = "43",
    number = "12",
    pages = "123001",
    year = "2026"
}

@article{Leung:1997was,
    author = "Leung, P. T. and Liu, Y. T. and Suen, W. M. and Tam, C. Y. and Young, K.",
    title = "{Quasinormal modes of dirty black holes}",
    eprint = "gr-qc/9903031",
    archivePrefix = "arXiv",
    doi = "10.1103/PhysRevLett.78.2894",
    journal = "Phys. Rev. Lett.",
    volume = "78",
    pages = "2894--2897",
    year = "1997"
}

@book{Chandrasekhar:1985kt,
    author = "Chandrasekhar, Subrahmanyan",
    title = "{The mathematical theory of black holes}",
    isbn = "978-0-19-850370-5",
    year = "1985"
}

@article{Li:2023ulk,
    author = "Li, Dongjun and Hussain, Asad and Wagle, Pratik and Chen, Yanbei and Yunes, Nicol{\'a}s and Zimmerman, Aaron",
    title = "{Isospectrality breaking in the Teukolsky formalism}",
    eprint = "2310.06033",
    archivePrefix = "arXiv",
    primaryClass = "gr-qc",
    reportNumber = "UTWI-14-2023",
    doi = "10.1103/PhysRevD.109.104026",
    journal = "Phys. Rev. D",
    volume = "109",
    number = "10",
    pages = "104026",
    year = "2024"
}

@article{Kehagias:2024rtz,
    author = "Kehagias, Alex and Riotto, Antonio",
    title = "{Black holes in a gravitational field: the non-linear static love number of Schwarzschild black holes vanishes}",
    eprint = "2410.11014",
    archivePrefix = "arXiv",
    primaryClass = "gr-qc",
    doi = "10.1088/1475-7516/2025/05/039",
    journal = "JCAP",
    volume = "05",
    pages = "039",
    year = "2025"
}

@article{Regge:1957td,
    author = "Regge, Tullio and Wheeler, John A.",
    title = "{Stability of a Schwarzschild singularity}",
    doi = "10.1103/PhysRev.108.1063",
    journal = "Phys. Rev.",
    volume = "108",
    pages = "1063--1069",
    year = "1957"
}

@article{Zerilli:1970se,
    author = "Zerilli, Frank J.",
    title = "{Effective potential for even parity Regge-Wheeler gravitational perturbation equations}",
    doi = "10.1103/PhysRevLett.24.737",
    journal = "Phys. Rev. Lett.",
    volume = "24",
    pages = "737--738",
    year = "1970"
}

@article{Lestingi:2026peq,
    author = "Lestingi, Jacopo and Sberna, Laura and Green, Stephen R.",
    title = {{Schr{\"o}dinger perturbation theory for black hole quasinormal modes}},
    eprint = "2607.19492",
    archivePrefix = "arXiv",
    primaryClass = "gr-qc",
    month = "7",
    year = "2026"
}

@article{Green:2022htq,
    author = "Green, Stephen R. and Hollands, Stefan and Sberna, Laura and Toomani, Vahid and Zimmerman, Peter",
    title = "{Conserved currents for a Kerr black hole and orthogonality of quasinormal modes}",
    eprint = "2210.15935",
    archivePrefix = "arXiv",
    primaryClass = "gr-qc",
    doi = "10.1103/PhysRevD.107.064030",
    journal = "Phys. Rev. D",
    volume = "107",
    number = "6",
    pages = "064030",
    year = "2023"
}

@article{Mark:2014aja,
    author = "Mark, Zachary and Yang, Huan and Zimmerman, Aaron and Chen, Yanbei",
    title = "{Quasinormal modes of weakly charged Kerr-Newman spacetimes}",
    eprint = "1409.5800",
    archivePrefix = "arXiv",
    primaryClass = "gr-qc",
    doi = "10.1103/PhysRevD.91.044025",
    journal = "Phys. Rev. D",
    volume = "91",
    number = "4",
    pages = "044025",
    year = "2015"
}

@article{papapetrou1953,
  title={Eine rotationssymmetrische L{\"o}sung in der allgemeinen Relativit{\"a}tstheorie},
  author={Papapetrou, Achille},
  journal={Annalen der Physik},
  volume={447},
  number={4-6},
  pages={309--315},
  year={1953},
  publisher={Wiley Online Library}
}

@article{Ernst:1967wx,
    author = "Ernst, Frederick J.",
    title = "{New formulation of the axially symmetric gravitational field problem}",
    doi = "10.1103/PhysRev.167.1175",
    journal = "Phys. Rev.",
    volume = "167",
    pages = "1175--1179",
    year = "1968"
}

@article{Binnington:2009bb,
    author = "Binnington, Taylor and Poisson, Eric",
    title = "{Relativistic theory of tidal Love numbers}",
    eprint = "0906.1366",
    archivePrefix = "arXiv",
    primaryClass = "gr-qc",
    doi = "10.1103/PhysRevD.80.084018",
    journal = "Phys. Rev. D",
    volume = "80",
    pages = "084018",
    year = "2009"
}

@article{Teukolsky:1973ha,
    author = "Teukolsky, Saul A.",
    title = "{Perturbations of a rotating black hole. 1. Fundamental equations for gravitational electromagnetic and neutrino field perturbations}",
    doi = "10.1086/152444",
    journal = "Astrophys. J.",
    volume = "185",
    pages = "635--647",
    year = "1973"
}

@article{Leaver:1985ax,
    author = "Leaver, E. W.",
    title = "{An Analytic representation for the quasi normal modes of Kerr black holes}",
    doi = "10.1098/rspa.1985.0119",
    journal = "Proc. Roy. Soc. Lond. A",
    volume = "402",
    pages = "285--298",
    year = "1985"
}

@article{DeLuca:2024uju,
    author = "De Luca, Valerio and Franciolini, Gabriele and Riotto, Antonio",
    title = "{Flea on the elephant: Tidal Love numbers in subsolar primordial black hole searches}",
    eprint = "2408.14207",
    archivePrefix = "arXiv",
    primaryClass = "gr-qc",
    reportNumber = "CERN-TH-2024-140",
    doi = "10.1103/PhysRevD.110.104041",
    journal = "Phys. Rev. D",
    volume = "110",
    number = "10",
    pages = "104041",
    year = "2024"
}

@article{Gounis:2024hcm,
    author = "Gounis, L. -R. and Kehagias, A. and Riotto, A.",
    title = "{The vanishing of the non-linear static love number of Kerr black holes and the role of symmetries}",
    eprint = "2412.08249",
    archivePrefix = "arXiv",
    primaryClass = "gr-qc",
    doi = "10.1088/1475-7516/2025/03/002",
    journal = "JCAP",
    volume = "03",
    pages = "002",
    year = "2025"
}

@book{Riotto:2024ayo,
    author = "Riotto, Antonio and Silk, Joe",
    title = "{The Future of~Primordial Black Holes: Open Questions and~Roadmap}",
    eprint = "2403.02907",
    archivePrefix = "arXiv",
    primaryClass = "astro-ph.CO",
    doi = "10.1007/978-981-97-8887-3-27",
    year = "2025"
}

@article{Maldacena:2015waa,
    author = "Maldacena, Juan and Shenker, Stephen H. and Stanford, Douglas",
    title = "{A bound on chaos}",
    eprint = "1503.01409",
    archivePrefix = "arXiv",
    primaryClass = "hep-th",
    doi = "10.1007/JHEP08(2016)106",
    journal = "JHEP",
    volume = "08",
    pages = "106",
    year = "2016"
}

@article{Cardoso:2008bp,
    author = "Cardoso, Vitor and Miranda, Alex S. and Berti, Emanuele and Witek, Helvi and Zanchin, Vilson T.",
    title = "{Geodesic stability, Lyapunov exponents and quasinormal modes}",
    eprint = "0812.1806",
    archivePrefix = "arXiv",
    primaryClass = "hep-th",
    doi = "10.1103/PhysRevD.79.064016",
    journal = "Phys. Rev. D",
    volume = "79",
    number = "6",
    pages = "064016",
    year = "2009"
}

@article{Yang:2012he,
    author = "Yang, Huan and Nichols, David A. and Zhang, Fan and Zimmerman, Aaron and Zhang, Zhongyang and Chen, Yanbei",
    title = "{Quasinormal-mode spectrum of Kerr black holes and its geometric interpretation}",
    eprint = "1207.4253",
    archivePrefix = "arXiv",
    primaryClass = "gr-qc",
    doi = "10.1103/PhysRevD.86.104006",
    journal = "Phys. Rev. D",
    volume = "86",
    pages = "104006",
    year = "2012"
}

@article{Fransen:2023eqj,
    author = "Fransen, Kwinten",
    title = "{Quasinormal modes from Penrose limits}",
    eprint = "2301.06999",
    archivePrefix = "arXiv",
    primaryClass = "gr-qc",
    doi = "10.1088/1361-6382/acf26d",
    journal = "Class. Quant. Grav.",
    volume = "40",
    number = "20",
    pages = "205004",
    year = "2023"
}

@article{Giataganas:2024hil,
    author = "Giataganas, D. and Kehagias, A. and Riotto, A.",
    title = "{Quasinormal modes and universality of the Penrose limit of black hole photon rings}",
    eprint = "2403.10605",
    archivePrefix = "arXiv",
    primaryClass = "gr-qc",
    doi = "10.1007/JHEP09(2024)168",
    journal = "JHEP",
    volume = "09",
    pages = "168",
    year = "2024"
}

@book{Penrose1976,
author="Penrose, Roger",
editor="Cahen, M.
and Flato, M.",
title="Any Space-Time has a Plane Wave as a Limit",
bookTitle="Differential Geometry and Relativity: A Volume in Honour of Andr{\'e} Lichnerowicz on His 60th Birthday",
year="1976",
publisher="Springer Netherlands",
address="Dordrecht",
pages="271--275",
isbn="978-94-010-1508-0",
doi="10.1007/978-94-010-1508-0\_23",
url="https://doi.org/10.1007/978-94-010-1508-0_23"
}

@article{Blau:2003dz,
    author = "Blau, Matthias and Borunda, Monica and O'Loughlin, Martin and Papadopoulos, George",
    title = "{Penrose limits and space-time singularities}",
    eprint = "hep-th/0312029",
    archivePrefix = "arXiv",
    reportNumber = "SISSA-102-2003-EP",
    doi = "10.1088/0264-9381/21/7/L02",
    journal = "Class. Quant. Grav.",
    volume = "21",
    pages = "L43",
    year = "2004"
}

@article{Jaramillo:2020tuu,
    author = "Jaramillo, Jos{\'e} Luis and Panosso Macedo, Rodrigo and Al Sheikh, Lamis",
    title = "{Pseudospectrum and Black Hole Quasinormal Mode Instability}",
    eprint = "2004.06434",
    archivePrefix = "arXiv",
    primaryClass = "gr-qc",
    doi = "10.1103/PhysRevX.11.031003",
    journal = "Phys. Rev. X",
    volume = "11",
    number = "3",
    pages = "031003",
    year = "2021"
}

@article{Cheung:2021bol,
    author = "Cheung, Mark Ho-Yeuk and Destounis, Kyriakos and Macedo, Rodrigo Panosso and Berti, Emanuele and Cardoso, Vitor",
    title = "{Destabilizing the Fundamental Mode of Black Holes: The Elephant and the Flea}",
    eprint = "2111.05415",
    archivePrefix = "arXiv",
    primaryClass = "gr-qc",
    doi = "10.1103/PhysRevLett.128.111103",
    journal = "Phys. Rev. Lett.",
    volume = "128",
    number = "11",
    pages = "111103",
    year = "2022"
}

@article{Graffi:1978gp,
  author = "Graffi, S. and Grecchi, V.",
  title = "{Resonances in Stark effect and perturbation theory}",
  journal = "Commun. Math. Phys.",
  volume = "62",
  pages = "83--96",
  year = "1978"
}

@article{Herbst:1979dil,
  author = "Herbst, I. W.",
  title = "{Dilation analyticity in constant electric field. I. The two body problem}",
  journal = "Commun. Math. Phys.",
  volume = "64",
  pages = "279--298",
  year = "1979"
}

@article{Cardoso:2021qqu,
    author = "Cardoso, Vitor and Foschi, Arianna",
    title = "{Geodesic structure and quasinormal modes of a tidally perturbed spacetime}",
    eprint = "2106.06551",
    archivePrefix = "arXiv",
    primaryClass = "gr-qc",
    doi = "10.1103/PhysRevD.104.024004",
    journal = "Phys. Rev. D",
    volume = "104",
    number = "2",
    pages = "024004",
    year = "2021"
}

@article{Bhowmik:2026owi,
    author = "Bhowmik, Arkadip and Chowdhury, Avijit and Chakrabarti, Sayan",
    title = "{The Ringdown and the Tide: Fingerprints of Dark Matter Halo Profiles}",
    eprint = "2608.07678",
    archivePrefix = "arXiv",
    primaryClass = "gr-qc",
    month = "8",
    year = "2026"
}

@article{Poisson:2014gka,
    author = "Poisson, Eric",
    title = "{Tidal deformation of a slowly rotating black hole}",
    eprint = "1411.4711",
    archivePrefix = "arXiv",
    primaryClass = "gr-qc",
    doi = "10.1103/PhysRevD.91.044004",
    journal = "Phys. Rev. D",
    volume = "91",
    number = "4",
    pages = "044004",
    year = "2015"
}

@article{Barausse:2014tra,
    author = "Barausse, Enrico and Cardoso, Vitor and Pani, Paolo",
    title = "{Can environmental effects spoil precision gravitational-wave astrophysics?}",
    eprint = "1404.7149",
    archivePrefix = "arXiv",
    primaryClass = "gr-qc",
    doi = "10.1103/PhysRevD.89.104059",
    journal = "Phys. Rev. D",
    volume = "89",
    number = "10",
    pages = "104059",
    year = "2014"
}

@article{Spieksma:2024voy,
    author = "Spieksma, Thomas F. M. and Cardoso, Vitor and Carullo, Gregorio and Della Rocca, Matteo and Duque, Francisco",
    title = "{Black Hole Spectroscopy in Environments: Detectability Prospects}",
    eprint = "2409.05950",
    archivePrefix = "arXiv",
    primaryClass = "gr-qc",
    doi = "10.1103/PhysRevLett.134.081402",
    journal = "Phys. Rev. Lett.",
    volume = "134",
    number = "8",
    pages = "081402",
    year = "2025"
}

@article{Zhao:2026eti,
    author = "Zhao, Yu-Qian and Pani, Paolo",
    title = "{Quasinormal modes and tidal responses of black holes in generic anisotropic matter environments}",
    eprint = "2606.11380",
    archivePrefix = "arXiv",
    primaryClass = "gr-qc",
    month = "6",
    year = "2026"
}

@article{Giataganas:2026fop,
    author = "Giataganas, D. and Giudice, G. F. and Kehagias, A. and Quevedo, F. and Riotto, A.",
    title = "{Thermal Origin of Black Hole Quasinormal Modes}",
    eprint = "2608.09797",
    archivePrefix = "arXiv",
    primaryClass = "hep-th",
    reportNumber = "CERN-TH-2026-189",
    month = "8",
    year = "2026"
}

@article{Giataganas:2026ctn,
    author = "Giataganas, D. and Giudice, G. F. and Kehagias, A. and Quevedo, F. and Riotto, A.",
    title = "{Black Hole Photon Rings Saturate the Quantum Chaos Bound}",
    eprint = "2605.29923",
    archivePrefix = "arXiv",
    primaryClass = "hep-th",
    reportNumber = "CERN-TH-2026-100",
    month = "5",
    year = "2026"
}

@article{Schutz:1985km,
    author = "Schutz, Bernard F. and Will, Clifford M.",
    title = "{Black Hole Normal Modes: A Semianalytic Approach}",
    reportNumber = "PRINT-85-0063 (WASH.U.,ST.LOUIS)",
    doi = "10.1086/184453",
    journal = "Astrophys. J. Lett.",
    volume = "291",
    pages = "L33--L36",
    year = "1985"
}

\end{document}